**Daniel Hopp**
Associate Statistician
Division on Globalisation and Development Strategies, UNCTAD
daniel.hopp@unctad.org

**Emily Fu**
Intern
Division on Globalisation and Development Strategies, UNCTAD
EmilyFu@cmail.carleton.ca

**Anu Peltola**
Acting Head Development Statistics and Information, UNCTAD
anu.peltola@unctad.org

# Feasibility of nowcasting SDG indicators: a comprehensive survey

## Abstract

The 2030 Agenda and accompanying Sustainable Development Goals (SDGs) are vital in guiding national and global policy. However, many of the SDG indicators used to measure progress toward those goals suffer from long publication lags. Nowcasting has the potential to address this problem and generate more timely estimates of those indicators. This paper provides resources for achieving that potential by 1) carrying out a comprehensive nowcasting feasibility survey of all SDG indicators to assess their potential to be nowcast, and 2) performing a case study of indicator 9.4.1 to illustrate and shed light on the process of performing a nowcasting exercise. There exist 231 SDG indicators, but due to only examining Tier 1 indicators and the fact that many indicators have multiple sub-indicators, 362 indicators and sub-indicators were eventually surveyed. Of those 362, 150 were found highly likely to be suitable candidates for nowcasting, 87 were found to be likely, and 125 were found to be unsuitable.

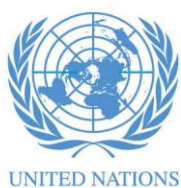

# Contents

## Acknowledgements

We would like to thank Bojan Nastav, Katalin Bokor, and Nour Barnat for their valuable comments and feedback.

# 1. Introduction

The Sustainable Development Goals (SDGs) of the 2030 Agenda were adopted by the United Nations General Assembly in 2015 in recognition of the need for an organized international framework to help address the myriad challenges facing the world in the 21$^{st}$ century (UN, 2015). The goals transformed considerably when compared with the 2000 Millennium Development Goals (MDGs), reflecting an increasing pace of technological, economic, and social change and applying to all countries globally instead of developing economies only. Some issues have remained timeless and appear in both the SDGs and MDGs, such as poverty, hunger, and education, while others were substantially expanded upon, such as those concerning the environment and climate change. Still many others, such as clean energy, were newly added.

Despite the ambitious aims of the 2030 Agenda, the UN recognized that its impact would be limited without proper means of measuring and quantifying progress on its goals. Consequently, the General Assembly asked the UN Statistical Commission to coordinate the substantive and technical work to develop the SDG indicator framework to measure targets selected for each goal, currently with a total of 231 indicators spread across the 17 goals (UNDESA, 2021a). While some indicators are similar or identical to existing statistics compiled and published by national statistical offices or other national authorities, such as unemployment rate, others needed to be newly defined and collected specifically for the 2030 Agenda. Furthermore, countries' data gaps vary greatly, and they have had to put in place special efforts to enable more comprehensive reporting on the indicator framework.

In addition to filling data gaps, national statistical authorities have been challenged by increasing pressure to provide more up-to-date information as evidence for policy makers so that they have enough time to influence progress towards achieving the goals of the Agenda by 2030. Poor timeliness is a common issue for many SDG indicators (UNSD, 2020). Indicators are of limited use to policy makers in terms of both planning and programming assessment if they are published with significant lags. As noted in the 'a World that Counts' report (United Nations, 2014), data delayed is data denied.

Recently, the rise of new technological possibilities and emerging digital data sources have enabled the compilation of timelier statistics. Numerous statistical offices have quickly responded to the demand for timely data during the COVID-19 pandemic, including with the use of non-traditional data sources and new statistical techniques. One such tool that could help address issues of timeliness in SDG indicators is nowcasting. Nowcasting is the estimation of the current, or near-current, value of a target series using information from more timely series. In a world awash in data from both a plethora of new sources and from new ways of storing old data (Einav and Levin, 2014), nowcasting can help leverage that information to obtain advance estimates of lower velocity indicators. As noted by MacFeely (2021), while nowcasting has generally been well received, many questions regarding the robustness of the methodologies employed need to be solved, as well as concerns over the validity of using a wide variety of data sources, including both hard and soft indicators. Concerns have also been flagged about the impact of revisions in the underlying data, dissemination strategies, potential confusion for users, division of work between international and national agencies, and relevance to some areas of sustainable development. To date, timeliness of SDG indicators has been the responsibility of the many separate SDG indicator custodian agencies, some managing better than others to improve timeliness in collaboration with national statistical authorities. Furthermore, as the indicators are mainly compiled by national statistical authorities, common approaches, methods and rules are needed. To increase collaboration on nowcasting among official statisticians, the United Nations

Conference on Trade and Development (UNCTAD) held a nowcasting workshop with United Nations Industrial Development Organization (UNIDO) in February 2020. The meeting discussed case studies on nowcasting exercises carried out in official statistics. In 2021, UNCTAD shared its experience on new nowcasting methodologies based on neural networks with the global statistical community at a UN Brown Bag seminar. This paper aims to help turn the potential to nowcast SDG indicators into reality by firstly providing a comprehensive survey of the nowcasting feasibility of all SDG indicators and secondly by fully documenting the process of nowcasting an SDG indicator via a case study. This work was carried out in the context of an informal 'nowcasting network', chaired by UNIDO and set up by UN Chief Statisticians.

For this paper, 362 SDG indicators and sub-indicators were surveyed for nowcasting feasibility. This number differs from the 231 mentioned above due firstly to only examining the 130 Tier 1 SDG indicators, and secondly to the fact that some indicators have several sub-indicators. See section 2.1 for more information on Tier 1 indicators. Of those 362, 150 received a classification of "Highly likely" able to be nowcast, 87 received a classification of "Likely", and 125 received a classification of "Unlikely". See sections 3.1 and 3.2 for more information on survey methodology and factors determining classification. Most indicators and sub-indicators were found to be recorded at an annual frequency, with publication lags ranging between one to three years, though these lags may differ by data source, which often varies by country or region of coverage. The existence of potential explanatory variables for use in modelling is unlikely to pose a problem for most indicators. Some indicators contain sets of sub-indicators surrounding a core subject matter. These sub-indicators typically share the same data availability and publication lags, with similar sets of likely explanatory variables.

The rest of the paper proceeds as follows: section 2 will provide further background on the SDG indicators and nowcasting; section 3 will describe the approach taken to complete the survey and report on general findings for each SDG; section 4 will present the empirical case study illustrating one approach to nowcasting an SDG indicator; section 5 will conclude, summarizing main results and recommendations going forward.

# 2. Background

## 2.1 SDG Indicators

In 2012, as the target year for the MDGs approached, work began on the development of a post-2015 development agenda (UNEP, 2012). The result was the 2030 Development Agenda, adopted by the UN General Assembly on the 25th of September, 2015 (UN, 2015). In contrast to the eight goals of the MDGs, the 2030 Agenda outlined 17 goals, called the Sustainable Development Goals (SDGs), which are accompanied by a varying number of targets per goal, for a total of 169 targets (UN, 2015). Each target in turn has one or more indicators to aid in monitoring progress towards accomplishment of each target and goal, for a total of 231 indicators (UNDESA, 2021a). Each of these indicators in turn has a varying number of sub-indicators. The 2030 Agenda serves as a policy framework to help tackle such issues as the eradication of poverty, reducing inequality, and addressing climate change, among many others. For more information on specific targets and indicators, see UNSD (2021a) and Ritchie et al. (2018).

Defining priorities and agreeing on a framework in 2015 was only part of the story, as a further two years were required to make the stated goals actionable by developing the targets and indicators for each goal (UN, 2017). Measurement is a vital aspect of the 2030 Agenda, both in terms of guiding and informing policy decisions at a national and international level, as well as in quantifying progress towards the goals. Developing indicators for such an expansive agenda with a diverse array of interconnections is no small task and highly dependent on the target and type of data available. As such, there are three tiers of SDG indicator:

- ***Tier 1***: *Indicator is conceptually clear, has an internationally established methodology and standards are available, and data are regularly produced by countries for at least 50 per cent of countries and of the population in every region where the indicator is relevant.*
- ***Tier 2***: *Indicator is conceptually clear, has an internationally established methodology and standards are available, but data are not regularly produced by countries.*
- ***Tier 3***: *No internationally established methodology or standards are yet available for the indicator, but methodology/standards are being (or will be) developed or tested* (UNSD, 2021b).

In simpler terms, Tier 1 indicators are well defined and already being produced, Tier 2 indicators are defined, but not yet being produced, while Tier 3 indicators are still being defined. It should also be noted that even indicators classified as Tier 1 may still be plagued by issues of data coverage, availability, and/or timeliness. Assessment of progress towards the achievement of the 2030 Agenda relies heavily upon reliable, accurate, disaggregated, and timely indicators. Nowcasting, examined in the next section, has the potential to address the last of those characteristics.

## 2.2 Nowcasting in the SDG context

The term "nowcast" itself is a portmanteau of "now" and "forecast". Nowcasting as a term and discipline originated in meteorology in the 1980s (WMO, 2017) but began to appear in economic literature in the 2000s. Nowcasting in the economic sense, and in the sense relevant for SDG indicators, refers to the estimation of the current value of a target variable based on timelier data and information. The distinction with forecasting comes from the fact that estimations are produced for time periods that either have already concluded or that are currently running, as opposed to periods in the future.

The intuition or justification for nowcasting is best explained by way of example using gross domestic product (GDP). GDP is a frequently nowcast target variable due to three of its characteristics. First, GDP is often published with a significant lag due to the many data sources needed and the complex accounting and aggregation procedures necessary for its calculation. GDP figures for a given quarter or year are often published many months after the conclusion of the period, even though all economic activity measured in the eventual figure has already occurred. Second, GDP usually has a long publication history, meaning there exist sufficient observations to estimate a model on historical information. Finally, there exist numerous potential explanatory variables which are published on a much timelier basis which can be used as inputs for a nowcast. Series such as consumer price indices, industrial production indices, consumer and business confidence indices, and retail trade figures, among many others, are typically published with a significantly shorter time lag than GDP, so can be used to obtain an estimate of GDP well before final figures are made available. These characteristics, together with the salience and relevance of GDP as an indicator, have made GDP the target variable for many nowcasting applications and papers. A GDP nowcasting model could then be fit on historical data and fed the latest information of timelier indicators to obtain both an estimate of GDP months before final figures are published, as well as monitor GDP outlook during the period by rerunning the model on the latest data continuously. For examples of GDP nowcasting applications see Morgado et al. (2007), Rossiter (2010), or Bok et al. (2018).

Nowcasting is relevant for SDG indicators because many face issues with timeliness. In order to successfully implement the 2030 Agenda for Sustainable Development, it is essential that policy makers have access to timely information as it relates to SDG indicators, a primary means of monitoring and evaluating progress and guiding policy interventions. The United Nations Statistics Division (UNSD) and UNCTAD have identified nowcasting as a key means of meeting this timeliness challenge (UNSD, 2020). The existing literature on nowcasting specifically as it relates to SDGs is sparse, but two notable works include Bierbaumer-Polly et al. (2019), where a comprehensive nowcasting exercise of SDG indicators using dynamic factor models is performed for Austria, and Hughes et al. (2021), where the International Futures forecasting system is used to nowcast many SDG-related indicators for more than 180 countries.

Nowcasting is however no panacea. It is only applicable for obtaining timelier estimates of an already produced SDG indicator. That restricts its application to Tier 1 indicators, where data are produced. For a given indicator to be suitable for nowcasting, a further two conditions need to be satisfied: the indicator needs a sufficiently long time series to be able to train a nowcasting model, and there need to exist sufficient related and timely explanatory variables. In order to assess the nowcasting feasibility of SDG indicators, these conditions were applied in carrying out the survey explained in greater detail in the next section. Complete survey results are available in Appendix 2, with a visualization of results available in Appendix 1.

Care should be taken if nowcasts are eventually adopted as advanced estimates of SDG indicators. Their status as data-based, quantitative estimates, liable to revision as the data outlook changes, should be clear to users, as well as when a figure has changed from a nowcast to its actual recorded value.

# 3. Nowcasting feasibility survey

## 3.1 Description of methodology

The first step in nowcasting an SDG indicator is determining whether it is even applicable to the case. That is, do the characteristics of the indicator fulfill the data requirements of nowcasting outlined in the previous section. This was the goal of this feasibility survey: to provide a comprehensive overview of every Tier 1 SDG indicator and their sub-indicators and their potential to be nowcast. The results of the survey could help statisticians and custodian agencies know at a glance whether their indicators have the potential to be nowcast and provide a springboard from which to launch their own investigations. There are no hard and fast rules for applying the earlier mentioned three conditions of nowcasting, which depend rather on the indicator. A one-month lag for an economic series may be considered a short lag, while for an epidemiological series it could be considered a long one, etc. Rather, each indicator needed to be examined individually and evaluated for nowcasting suitability on a holistic basis. It is also worth mentioning that some indicators may be composed of a combination of two or more series, for instance indicator 8.4.2 (*Domestic material consumption, domestic material consumption per capita, and domestic material consumption per GDP*). In these cases, the indicator's publication delay may be due to one of its constituent series, and better results may be obtained from nowcasting just this series rather than the entire indicator itself.

As Tier 2 and 3 indicators lack the publication of any historical data, they could immediately be classified as not suitable for nowcasting due to the second condition. As such, they were excluded from the survey and do not appear in the table in Appendix 2. Restricted to Tier 1 indicators, the survey was conducted in the following manner: the main sources for information on Tier 1 indicators were the SDG Indicators Metadata Repository and the Global SDG Indicator Database (UNSD, 2021a; UNDESA, 2021b). The SDG Indicators Metadata Repository ideally includes information on data characteristics relevant to nowcasting feasibility. However, the type of information included in each SDG indicator metadata file tends to vary, despite ongoing work to standardize the contents. For instance, some files do not include any information on when the data are collected or released. Some metadata files are unfinished, have missing parts, or require updating or reviewing. The SDG Indicator Database displays data for each indicator but does not always reflect the data availability described for each indicator in the respective metadata file.

Custodian agency databases also usually provide access to data for their SDG indicators. Information from these databases was used as a direct source of indicator characteristics or used to validate metadata. Data availability in the SDG Indicator Database is usually up to date with the custodian agency databases but may also be vastly different in content. The survey combines data available in the two sources if the years do not completely overlap. Otherwise, the source with the longest time series provides the information in Appendix 2. The SDG Indicator Database is also limited to displaying annual data. Indicators with monthly or quarterly data, for instance, are only displayed as annual. Other database sources must be used to get information about these indicators. Data availability is not easily describable for some indicators and sub-indicators. In many cases the length of time series varies greatly by country or region. This is indicated in Appendix 2.

For an indicator to have Tier 1 classification, data must be available for over 50% of relevant countries. There are a few indicators that can variably be classified as Tier 1 or Tier 2, depending on sub-indicator, and some sub-indicators included appear to not meet the Tier 1 requirement stated above. The number of countries or territories covered by each indicator across all possible sources is difficult to confirm due to inaccurate metadata information and inconsistencies between data sources. Many SDG indicators are also not classified at the sub-indicator level, or existing classifications have changed over time without metadata updates. Some are broadly considered Tier 1 indicators despite having sub-indicators that may not meet the Tier 1 requirements, while others are classified as either Tier 1 or Tier 2 by sub-indicator, but without specifying which sub-indicators belong to which classification.

The amount of data available for a particular indicator can vary by country, location, or aggregate grouping, so the data availability described in Appendix 2 generally focused on the data availability of a world or global aggregate, if available. When the global aggregate did not have enough data for nowcasting purposes or a global aggregate was not available for a particular indicator, for the purposes of this survey, a general summary of the relevant countries and other aggregations was used. Information on publication lags for each indicator often had to be inferred from the data that appeared to be available and the existing metadata information.

A further consideration for nowcasting feasibility is the existence of explanatory variables for a particular indicator. Each indicator or sub-indicator was given a score of "Highly likely," "Likely," or "Unlikely" for this area. An indicator got a label of "Highly likely" if explanatory variables would likely be easy to find. Many SDG indicators are macroeconomic variables, such as GDP, or fall under poverty, health, education, environment or ecological topics. Variables like these are frequently modelled and existence and availability of explanatory variables is well-documented in literature. An indicator gets a label of "Likely" if potential explanatory variables may not be closely related to the behavior of the indicator or had limited data available. For instance, variables related to Official Development Assistance (ODA) or certain government spending decisions were listed in this category, since values for these indicators are generally pre-determined by government decisions, although there are a variety of socioeconomic factors that may still contribute to initial spending decisions themselves. An indicator got a label of "Unlikely" if it may prove difficult to find explanatory variables. Appropriate explanatory variables may not exist for binary outcome variables that capture whether a country or territory has enacted a certain policy or legislation or joined a certain agreement or organization, as these decisions are generally not reversed or changed once decided upon. A regional or global aggregate of such an indicator, e.g., disclosing the number of countries or economies in a region adhering to a particular policy, legislation or agreement, etc., would be more suitable, if it exists. It should be noted that classifications for existence of explanatory variables in the survey should serve only as a starting point. Definitively determining whether explanatory variables exist and selecting them for an SDG indicator can only be done with extensive research and potentially modelling, which was not feasible for this survey due to the quantity of SDG indicators and time and resource constraints.

Finally, scores for overall nowcasting feasibility were determined by all gathered information on data availability, publication lags, and explanatory variables. Each SDG indicator and sub-indicator was given a score of "Highly likely," "Likely," or "Unlikely" for overall feasibility. Some indicators were determined to be less feasible for nowcasting due to the nature of the subject matter being unsuitable for nowcasting purposes, such as the previously mentioned binary outcome indicators and indicators related to spending and budgetary decisions, as well as election result indicators or upper parliament appointment results. Some indicators were considered unsuitable for nowcasting if they did not have enough data, generally around 10 data points at a minimum. Finally, if an indicator is published without significant data lags, nowcasting may not be relevant for the case. As mentioned previously, what constitutes a significant lag depends on the indicator.

## 3.2 Survey results

Out of a total of 362 Tier 1 indicators and sub-indicators considered in the nowcasting feasibility survey, 150 received a classification of "Highly likely" for nowcasting feasibility, 87 received a classification of "Likely", and 125 received a classification of "Unlikely." The release of data for nearly all Tier 1 indicators and sub-indicators are accompanied by a lag. Most indicators are published at an annual frequency, with a publication lag of around one to three years. Publication lags may also differ by the organization, country, or region that provides the data. In these cases, the compilation of aggregate figures may depend on the "last" country or region to provide indicator estimates, so data publication lags are usually sufficient to warrant a nowcasting approach for both individual and aggregate figures alike. The existence of explanatory variables is unlikely to pose a problem for most indicators.

Some sub-indicators under the same indicator code measure identical concepts with different units, such as "Number of deaths and missing persons attributed to disasters" and "Number of deaths and missing persons attributed to disasters per 100,000 population" under Indicator 1.5.1. Usually, sub-indicators under the same indicator code share the same nowcasting feasibility as they share the same data availability, data release schedule, and generally surround related topics with similar sets of likely explanatory variables.

The following sections will summarize the survey results by goal, considering only Tier 1 indicators. For more detailed information on the results for a particular goal, indicator, or sub-indicator, see the full results table in Appendix 2.

### Goal 1: No Poverty

Goal 1 contains 15 Tier 1 indicators and sub-indicators. Feasibility information for this goal was sourced from the SDG indicator database and the SDG indicator metadata repository. Of the 15 indicators, five were found to be "Highly likely" and ten were found to be "Likely" suitable for nowcasting. Importantly, all indicators were found to have at least ten years of annual data available, with publication lags sufficiently long enough to warrant consideration for nowcasting. In terms of explanatory variables, much work already exists on nowcasting poverty and poverty-related indicators at the national and regional level (Aguilar et al., 2019; Makdissi, 2021; Navicke et al., 2014; Browne et al., 2021). As such, suitable, timely explanatory variables should exist. For instance, the World Bank's World Development Indicators (WDI) database contains numerous series which could be used in nowcasting goal 1 indicators..

Specific observations include the fact that indicator 1.4.1's (*proportion of population living in households with access to basic services*) release schedule appears inconsistent from year to year, with data released every three to five years. Additionally for indicator 1.5.1 (*number of deaths, missing persons and directly affected persons attributed to disasters per 100,000 population*), while natural disasters themselves may be difficult to predict, their economic and human impacts, the focus of indicator 1.5.1, remain feasible for nowcasting.

## Goal 2: Zero Hunger

Goal 2 contains 25 Tier 1 indicators and sub-indicators covering topics including incidence of health-related diagnoses, agriculture, and food prices. Feasibility information for this goal was sourced from the SDG indicator database, the SDG indicator metadata repository, and the Food and Agriculture Organization of the United Nations (FAO). Of the 25 indicators, 14 were found to be "Highly likely" suitable for nowcasting, three were found to be "Likely", and eight were found to be "Unlikely". Most indicators were found to have over ten years of observations while indicators labelled "Unlikely" all had insufficient observations for nowcasting purposes. Regarding potential explanatory variables, generally there is much timely and quality health data available for training a nowcasting model, as these topics are frequently modelled to assess policy impacts and determine strategies for management of health phenomena (Browne et al., 2021; FAO et al., 2021; Kim et al., 2017).

Similar to indicator 1.4.1, data for sub-indicators of 2.2.3 (*prevalence of anaemia in women aged 15 to 49 years, by pregnancy status*) are not released on a consistent basis from year to year, being released every three to five years. Nowcasting may still be beneficial for these sub-indicators as there will always be some publication lag. While most indicators have over one decade of annual data, the 2.c.1 sub-indicator (*consumer food price index*) has annual and monthly data available depending on the source used to access the data.

## Goal 3: Good Health and Well-Being:

Goal 3 contains 45 Tier 1 indicators and sub-indicators, covering topics including birth and death rates, incidence of health-related diagnoses, and access to health facilities. Feasibility information for this goal was sourced from the SDG indicator database, the SDG indicator metadata repository, and the World Health Organization (WHO). Of the 45 indicators, 18 were found to be "Highly likely" suitable for nowcasting, seven were found to be "Likely", and 20 were found to be "Unlikely". Explanatory variables for a nowcasting model around health indicators are likely to exist, given the extent to which they are already modelled and forecasted (van de Kassteele et al., 2019; Spreco et al., 2018; Nsoesie et al., 2020; Johansson et al., 2014; Wu et al., 2020). The primary reason for a Goal 3 indicator to get a label of "Unlikely" suitable for nowcasting was data availability. Many indicators did not have a suitably long time series, for instance with either only a single data point or data only every five years. Some indicators were classified as "Likely" if there was sparse data availability at the country or aggregate level. Nowcasting for specific countries or regions remains feasible for these cases.

Indicators 3.8.1 (*coverage of essential health services*), 3.b.2 (*total net official development assistance to medical research and basic health sectors*), and sub-indicator "coverage of treatment interventions for substance use disorders" of indicator 3.5.1 were classified as "Likely" as they could be considered accounting or budget-type indicators.

## Goal 4: Quality Education

Goal 4 contains 26 Tier 1 indicators and sub-indicators, covering topics including education completion rates, parity indices, and school resource access. Feasibility information for this goal was sourced from the Open SDG Data Hub, the United Nations

Educational, Scientific and Cultural Organization (UNESCO) Institute for Statistics (UIS), SDG indicator database, and SDG indicator metadata. Of the 26 indicators, one was found to be "Highly likely" suitable for nowcasting, 15 were found to be "Likely", and ten were found to be "Unlikely". Most Goal 4 indicators are sparsely reported at the country or regional level, but data for some individual countries and regions may have sufficiently long publication histories for nowcasting purposes. Generally, data publication for the Goal 4 indicators varies by region. Explanatory variables for education-related indicators are likely to be widely available, as there are a variety of socioeconomic factors that impact students and school systems. Modelling is a common approach for analyzing various metrics of education (Friedman et al., 2020).

## Goal 5: Gender Equality

Goal 5 contains eight Tier 1 indicators and sub-indicators, covering topics including child marriage, government, and employment. Feasibility information for this goal was sourced from the Inter-Parliamentary union (IPU), the UN Economic Commission for Europe (UNECE), the International Labour Organization (ILO), the SDG indicator database, and SDG indicator metadata. Of the eight indicators, two were found to be "Highly likely" suitable for nowcasting, four were found to be "Likely", and two "Unlikely". Explanatory variables for these indicators are generally likely to exist, as models for measures of gender equality are commonly used to track and forecast equality progress (Doorley et al., 2021; Chen, 2020; Rodríguez-Rodríguez et al., 2020; Ahrens and van der Vleuten, 2020; Fatehkia et al., 2018), as well as measure the impacts of economic shocks, policies, and other events.

The 5.5.1 sub-indicators "number of seats held by women in national parliaments", "proportion of elected seats held by women in deliberative bodies of local government", and "proportion of seats held by women in national parliaments" are given the feasibility classification of "Likely", as the makeups of local and national parliaments generally only change after elections take place, and results of individual elections are usually published without significant lags. It may be more suitable to look at specific upcoming elections individually. The 5.5.1 sub-indicator "current number of seats in national parliaments" is likely unsuitable for nowcasting as it doesn't generally change over time.

## Goal 6: Clean Water and Sanitation

Goal 6 contains 41 Tier 1 indicators and sub-indicators covering topics including water law and policy, water use, and water area. Feasibility information for this goal was sourced from the WHO, the UN Environment Programme (UNEP), the SDG indicator database, and SDG indicator metadata. Of the 41 indicators, 27 were found to be "Highly likely" suitable for nowcasting, one was found to be "Likely", and 13 were found to be "Unlikely". Over half of the Goal 6 indicators come from 6.6.1's sub-indicators. In general, Goal 6 includes indicators that fall under ecological or environmental topics, for which modelling is frequently used (Searcy et al., 2018; Dada, 2019; Seed et al., 2013; Proisy et al., 2016; Bouget et al., 2021; Reimer and Wu, 2016). Explanatory variables should be widely available for these indicators. Most indicators have around a two-year publication lag, with either several decades or only a few years of annual data available.

Similar to Indicator 1.4.1, data for Indicator 6.5.1 (*degree of integrated water resources management*) is not released on a consistent basis year to year, being released every 3 to 5 years. Indicator 6.a.1 (*amount of water- and sanitation-related official development assistance that is part of a government-coordinated spending plan*) was classified as "Likely" as it could be considered an accounting or budget-type indicator. Indicator 6.b.1 (*proportion of local administrative units with established and operational policies and procedures for participation of local communities in water and sanitation management*) covers laws/policies relating to water, which are likely not suitable for nowcasting due to both insufficient data availability and lack of aggregate reporting.

## Goal 7: Affordable and Clean Energy

Goal 7 contains six Tier 1 indicators and sub-indicators, covering topics including renewable energy, electricity access, and energy intensity. Feasibility information for this goal was sourced from the SDG indicator database and SDG indicator metadata. Of the 6 indicators, five were found to be "Highly likely" suitable for nowcasting, one was found to be "Likely", and none were found to be "Unlikely". All indicators/sub-indicators have a one to two-year publication lag and over a decade of annual data available. In general, Goal 7 includes indicators that fall under ecological or environmental topics, for which modelling is frequently used (Cheng, 2017; Samu et al., 2020). Explanatory variables should be widely available for these indicators.

## Goal 8: Decent Work and Economic Growth

Goal 8 contains 13 Tier 1 indicators and sub-indicators, covering topics including macroeconomic variables, commercial banks, and Aid for Trade. Feasibility information for this goal was sourced from the ILO, the SDG indicator database and SDG indicator metadata. Of the 13 indicators, seven were found to be "Highly likely" suitable for nowcasting, five were found to be "Likely", and one was found to be "Unlikely". Nowcasting is often used to analyze macroeconomic variables like those included in the Goal 8 indicators (Siliverstovs, 2020; Marcellino and Schumacher, 2010; Pavlicek and Kristoufek, 2015; Caperna et al., 2020). Explanatory variables should be widely available for these indicators.

Data for the 8.4.2 (*Domestic material consumption, domestic material consumption per capita, and domestic material consumption per GDP*) sub-indicators has not been released since 2017, and publication lags are unknown. Nowcasting may be especially suitable for these sub-indicators given the lack of more recent data. Data for Indicators 8.5.2 (*unemployment rate, by sex, age and persons with disabilities*), 8.6.1 (*proportion of youth (aged 15–24 years) not in education, employment or training*), and 8.10.1 (*number of commercial bank branches per 100,000 adults and (b) number of automated teller machines (ATMs) per 100,000 adults*) are collected by individual financial regulators or statistical organizations. Data release for these indicators varies by individual data source. The 8.a.1 (*Aid for Trade commitments and disbursements*) sub-indicators were classified as "Likely" as they may be considered accounting or budget-type indicators. Indicator 8.10.2 (*proportion of adults (15 years and older) with an account at a bank or other financial institution or with a mobile-money-service provider*) is the only Goal 8 indicator labelled as "Unlikely" for nowcasting feasibility, as it did not have a suitably long time series, with only three data points available.

## Goal 9: Industry, Innovation and Infrastructure

Goal 9 contains 19 Tier 1 indicators and sub-indicators, covering topics including macroeconomic variables, telecommunications, and carbon dioxide emissions. Feasibility information for this goal was sourced from UNCTAD, UNESCO, the Organisation for Economic Co-operation and Development (OECD), the SDG indicator database and SDG indicator metadata. Of the 19 indicators, 16 were found to be "Highly likely" suitable for nowcasting, one was found to be "Likely", and two were found to be "Unlikely". Nowcasting is often used to analyze macroeconomic variables like those included in the Goal 9 indicators (Sanyal and Das, 2018; Boudt et al., 2009; Hussain et al., 2018; Wiebe and Yamano, 2015; Arslanalp et al., 2019). Explanatory variables should be widely available for these indicators. Indicator 9.4.1 (*CO2 emission per unit of value added*) is considered in detail in Section 4 to demonstrate the process of selecting and performing a modelling exercise on a feasible SDG indicator.

The sub-indicators under Indicator 9.1.2 (*passenger and freight volumes, by mode of transport*) are the only Goal 9 indicators given a score of "Unlikely", as they do not have suitably long publication histories. Data publication lags for Indicators 9.2.2 (*manufacturing employment as a proportion of total employment*) and 9.3.2 (*proportion of small-scale industries with a loan or line of credit*) vary by data source, but both aggregate values and values for select countries and regions are likely good candidates for nowcasting. Indicator 9.a.1 (*total official international support (official development assistance plus other official flows) to infrastructure*) is classified as "Likely" as it may be considered an accounting or budget-type indicator.

## Goal 10: Reduced Inequalities

Goal 10 contains 19 Tier 1 indicators and sub-indicators, covering topics including financial markets, resource flows, developing countries, and refugees. Feasibility information for this goal was sourced from the Missing Migrants Project, the World Bank, the SDG indicator database and SDG indicator metadata. Of the 19 indicators, 12 were found to be "Highly likely" suitable for nowcasting, four were found to be "Likely", and three were found to be "Unlikely". Publication lags for 11 indicators and sub-indicators vary by data source.

There are multiple possible sources from which to access data on Indicator 10.7.3 (*number of people who died or disappeared in the process of migration towards an international destination*) that differ in periodicity and publication lag, with annual data released on the SDG indicator database and data at the incident level from the Missing Migrants Project. Although annual data has around a one-year lag, nowcasting may not be useful if more recent data is available as incidents happen. The 10.b.1 (*total resource flows for development, by recipient and donor countries and type of flow*) sub-indicators are classified as "Likely" as they may be considered accounting or budget-type indicators. Sub-indicators under 10.c.1 (*remittance costs as a proportion of the amount remitted*) are reported quarterly, with over 4 years of quarterly data. All other indicators and sub-indicators aside from 10.7.3 (*number of people who died or disappeared in the process of migration towards an international destination*) and 10.c.1 are annual, with 12 having over ten years of annual data.

## Goal 11: Sustainable Cities and Communities

Goal 11 contains 13 Tier 1 indicators and sub-indicators, covering topics including natural disasters and living conditions. Feasibility information for this goal was sourced from the SDG indicator database and SDG indicator metadata. Of the 13 indicators, none were found to be "Highly likely" suitable for nowcasting, ten were found to be "Likely", and three were found to be "Unlikely". In general, Goal 11 includes indicators that fall under ecological or environmental topics, for which modelling is frequently used (Kosmopoulos et al., 2018; Yu et al., 2015). Explanatory variables should be widely available for these indicators.

Data publication lags for indicator 11.1.1 (*proportion of urban population living in slums, informal settlements or inadequate housing*) varies by individual data source. The 11.5.1 (*number of deaths, missing persons and directly affected persons attributed to disasters per 100,000 population*) natural disaster sub-indicators are identical to the 1.5.1 and 13.1.1 natural disaster sub-indicators. These are considered "Likely" for nowcasting feasibility. As noted for the 1.5.1 sub-indicators, while natural disasters themselves may be difficult to predict, their economic and human impacts remain feasible for nowcasting. Aside from the natural disaster sub-indicators, the only unique Tier 1 indicators under Goal 11 are Indicators 11.1.1, 11.6.2 (*annual mean levels of fine particulate matter (e.g. PM2.5 and PM10) in cities (population weighted)*), and 11.a.1 (*number of countries that adopt and implement national disaster risk reduction strategies in line with the Sendai Framework for Disaster Risk Reduction 2015–2030*), which do not have suitably long time series for nowcasting purposes.

## Goal 12: Responsible Consumption and Production

Goal 12 contains 15 Tier 1 indicators and sub-indicators, covering topics including environmental sustainability and domestic material consumption. Feasibility information for this goal was sourced from the SDG indicator database and SDG indicator metadata. Of the 15 indicators, four were found to be "Highly likely" suitable for nowcasting, three were found to be "Likely", and eight were found to be "Unlikely". In general, Goal 12 includes indicators that fall under ecological or environmental topics, for which modelling is frequently used (Bennedsen et al., 2021; Videnova et al., 2006). Explanatory variables should be widely available for these indicators.

The 12.2.2 (*domestic material consumption, domestic material consumption per capita, and domestic material consumption per GDP*) sub-indicators are identical to the 8.4.2 domestic material consumption sub-indicators and are also considered "Highly likely" for nowcasting feasibility. As with the 8.4.2 sub-indicators, data has not been released since 2017, and publication lags are unknown. Nowcasting may be especially suitable for these sub-indicators given the lack of more recent data. All 12.4.1 (*number of parties to international multilateral environmental agreements on hazardous waste, and other chemicals that meet their commitments and obligations in transmitting information as required by each relevant agreement*) sub-indicators have only two years of data available, so are not feasible for nowcasting.

## Goal 13: Climate Action

Goal 13 contains 12 Tier 1 indicators and sub-indicators, covering topics including natural disasters and greenhouse gas emissions. Feasibility information for this goal was sourced from the SDG indicator database and SDG indicator metadata. Of the 12 indicators, all were found to be "Highly likely" suitable for nowcasting, the only SDG goal for which all Tier 1 indicators have this classification. All indicators have a publication lag of one to three years and over ten years of annual data. In general, Goal 13 includes indicators that fall under ecological or environmental topics, for which modelling is frequently used (Bennedsen et al., 2021; Lamboll et al., 2021; Wiebe and Yamano, 2015). Explanatory variables should be widely available for these indicators.

The 13.1.1 (*number of deaths, missing persons and directly affected persons attributed to disasters per 100,000 population*) natural disaster sub-indicators are identical to the 1.5.1 and 11.5.1 natural disaster sub-indicators. These are considered "Likely" for nowcasting feasibility. As noted for the 1.5.1 sub-indicators, while natural disasters themselves may be difficult to predict, their economic and human impacts remain feasible for nowcasting. Data availability for the 13.2.2 sub-indicator "Total greenhouse gas emissions without LULUCF for non-Annex I Parties" varies widely by region and no aggregations are produced. However, there are many countries with sufficient data for nowcasting purposes.

## Goal 14: Life Below Water

Goal 14 contains seven Tier 1 indicators and sub-indicators, covering topics including fishing, marine area use, and fishing law and policy. Feasibility information for this goal was sourced from the SDG indicator database and SDG indicator metadata. Of the seven indicators, four were found to be "Highly likely" suitable for nowcasting, none were found to be "Likely", and three were found to be "Unlikely". In general, Goal 14 includes indicators that fall under ecological or environmental topics, for which modelling is frequently used (Hobday et al., 2011; Kohut et al., 2012; Carter et al., 2015). Explanatory variables should be widely available for these indicators.

Indicator 14.6.1 (*degree of implementation of international instruments aiming to combat illegal, unreported and unregulated fishing*) is unsuitable for nowcasting, with only two years of available data. Additionally, data for indicators 14.7.1 (*sustainable fisheries as a proportion of GDP in small island developing States, least developed countries and all countries*) and 14.b.1 (*degree of application of a legal/regulatory/ policy/institutional framework which recognizes and protects access rights for small-scale fisheries*) have biennial periodicity and biennial data release. They do not have sufficient observations for nowcasting purposes and are thus labelled "Unlikely" for nowcasting feasibility.

## Goal 15: Life on Land

Goal 15 contains 32 Tier 1 indicators and sub-indicators, covering topics including forests, endangered species, environmental law and policy, and ODA. Feasibility information for this goal was sourced from the SDG indicator database and SDG indicator metadata. Of the 32 indicators, five were found to be "Highly likely" suitable for nowcasting, four were found to be "Likely", and 23 were found to be "Unlikely". In general, Goal 15 includes indicators that fall under ecological or environmental topics, for which modelling is frequently used (Staver and Levin, 2012; Bugmann et al., 2010). Explanatory variables should be widely available for these indicators. Notably, over half of the Goal 15 indicators and sub-indicators are likely unsuitable for nowcasting due to insufficient observations.

For the 15.1.1 sub-indicators "Forest area" and "Forest area as a proportion of total land area", it was unclear if there is a usual pattern to data release and if there is a consistent publication lag. The 15.a.1 and 15.b.1 (*official development assistance on conservation and sustainable use of biodiversity; and (b) revenue generated and finance mobilized from biodiversity-relevant economic instruments*) sub-indicators are classified as "Likely" as they may be considered accounting or budget-type indicators.

## Goal 16: Peace, Justice and Strong Institutions

Goal 16 contains 25 indicators and sub-indicators, covering topics including crime, governments, elections, and the Paris Principles. Feasibility information for this goal was sourced from the Inter-Parliamentary Union (IPU), the SDG indicator database and SDG indicator metadata. Of the 25 indicators, one was found to be "Highly likely" suitable for nowcasting, eight were found to be "Likely", and 16 were found to be "Unlikely". Lack of publication history is a primary factor in the unsuitability of many Goal 16 indicators.

Data availability for Indicator 16.5.2 (*proportion of businesses that had at least one contact with a public official and that paid a bribe to a public official, or were asked for a bribe by those public officials during the previous 12 months*) varies by individual region with generally sporadic data publishing, but sufficient data exist for select countries/regions or aggregates. Because elections data are released and information on parliaments is updated relatively quickly, nowcasting may be less applicable to the 16.7.1 (*proportions of positions in national and local institutions, including (a) the legislatures; (b) the public service; and (c) the judiciary, compared to national distributions, by sex, age, persons with disabilities and population groups*) sub-indicators. However, modelling election results for specific countries or regions is common. Upper chamber parliaments generally do not have elected positions, but there still may be broad social or economic factors that impact appointments to such positions as well as elections results as a whole. Annual data for these sub-indicators is updated annually for the current year, so publication lags will depend on election dates falling before or after data release. More recent data can be found by following country elections individually. Data publication for the 16.8.1 (*proportion of members and voting rights of developing countries in international organizations*) sub-indicators and Indicator 16.9.1 (*proportion of children under 5 years of age whose births have been registered with a civil authority, by age*) depend on the region or organization supplying the data. Nowcasting may be less applicable to Indicator 16.10.1 (*number of verified cases of killing, kidnapping, enforced disappearance, arbitrary detention and torture of journalists, associated media personnel, trade unionists and human rights advocates in the previous 12 months*) as it is published without an annual lag. It also only has one year of available data, so would be unsuitable for nowcasting given current time series availability.

## Goal 17: Partnerships for the Goals

Goal 17 contains 41 indicators and sub-indicators, covering topics including financial assistance to developing countries and macroeconomic variables. Feasibility information for this goal was sourced from the Instituto Nacional de Estadística (INE) of Spain, the OECD, the SDG indicator database and SDG indicator metadata. Of the 41 indicators, 13 were found to be "Highly likely" suitable for nowcasting, 15 were found to be "Likely", and 13 were found to be "Unlikely". Nowcasting is often used to analyze macroeconomic variables like those included in the Goal 17 indicators (Lahiri and Yang, 2021; Strohsal and Wolf, 2020; Chapman and Desai, 2021; Jafari and Britz, 2018; Himics et al., 2020; Bekkers et al., 2020). Explanatory variables should be widely available for these indicators.

The 17.2.1 (*net official development assistance, total and to least developed countries, as a proportion of the Organization for Economic Cooperation and Development (OECD) Development Assistance Committee donors' gross national income (GNI)*) ODA sub-indicators were classified as "Likely" as they may be considered accounting or budget-type indicators. The two 17.6.1 (*fixed Internet broadband subscriptions per 100 inhabitants, by speed*) sub-indicators and Indicator 17.8.1 (*proportion of individuals using the Internet*) are related to global internet access. Data release for these indicators varies at the country or regional level. Indicator 17.18.2 (*number of countries that have national statistical legislation that complies with the Fundamental Principles of Official Statistics*), the 17.18.3 (*number of countries with a national statistical plan that is fully funded and under implementation, by source of funding*) sub-indicators, and the sub-indicators "countries that have conducted at least one population and housing census in the last 10 years," "countries with birth registration data that are at least 90 percent complete," and "countries with death registration data that are at least 75 percent complete" under 17.19.2 are all binary variables that have insufficient data publication histories for nowcasting purposes. 2020 data is the most recently available data for the 17.18.3 sub-indicators, but there is insufficient information to determine data release schedules and usual publication lags, if they exist.

# 4. Empirical Pilot

## 4.1 Indicator and data

The case study presented here will demonstrate how the information from the feasibility survey can be used to select promising SDG indicators for nowcasting and provide guidance on how the modelling exercise can then be performed. The selected indicator for the exercise was indicator 9.4.1, $CO_2$ emissions per unit of value added (UNSD, 2021c), on the global level. As with most indicators, there exist many different regional and country-level aggregations, each with their own publication schedules and data availability characteristics. The global level was chosen as the aggregation of broadest interest. Examining the survey results in Appendix 2, we can see that 9.4.1 has data from 2000 onwards, thus satisfying the adequate series history requirement of nowcasting, is published on the SDG database with a multi-year lag, satisfying the requirement of an extended publication lag, and that there exist ample potential explanatory variables. These characteristics together make it a good candidate for nowcasting and for the case study.

Within the SDG indicator, the series *EN_ATM_CO2GDP*, "carbon dioxide emissions per unit of GDP (kilogrammes of $CO_2$ per constant 2017 United States dollars)" was used as the nowcasting target variable (UNDESA, 2021b). Data for this series were obtained from the UN SDG database (UNDESA, 2021b), where, at the time of writing in Autumn 2021, the latest global figures were available for 2000-2018 at an annual frequency. Data for the database in turn come from the International Energy Agency (IEA) (IEA, 2021). It should be noted that while data available on the SDG database would imply a publication lag of more than two and a half years, timelier data may be available from the IEA directly. However, at the time of writing, timelier figures for the indicator are not available publicly. Any custodian agency, institution, or individual interested in nowcasting an SDG indicator should first make sure that timelier data are not available directly from the original data provider. For the purposes of this case study, where the goal is not to generate a nowcast of the indicator per se, but rather to illustrate and outline the modelling process, we can take the publicly available publication lag of two years as given.

With data for the target indicator in hand, the next step in the nowcasting process is identifying and gathering data for potential explanatory variables. The actual variables identified depend highly on which indicator is being nowcast. It is recommended to gather as many potentially related variables as possible. A process for selecting which variables go into the model in the end will be outlined in the next section. In the case of 9.4.1, there were two main components of the indicator: carbon emissions and economic activity, i.e., GDP. The data gathering process could then be guided by these two components.

Data on emissions were mainly gathered from two sources, Statista and the U.S. Energy Information Administration (Statista, 2021; EIA, 2021). Data from the former had a publication lag of between four and eight months on annual-frequency data, while data from the latter had a publication lag of three months on monthly-frequency data. Data relating to GDP and economic activity were drawn from many sources, but mainly from the OECD (OECD, 2021) and Eurostat (Eurostat, 2021). Economic data had any of a monthly, quarterly, or yearly frequency. In the end, almost 30 emissions-related and more than 150 economy-related variables were gathered. Not all these variables were used in the final model, they rather served as a pool from which to train and test different models in the model selection process outlined in the next section. All series were finally transformed to period over period seasonally adjusted (if applicable) growth rates with the US Census Bureau's X13-ARIMA-SEATS methodology (USCB, 2017).

## 4.2 Methodology

Once the data have been gathered, three steps remain in the modeling process: selecting a modelling methodology, selecting which variables will go into the model, and selecting which hyperparameters to use for the chosen methodology. The last step depends on the type of model chosen, as some approaches do not have hyperparameters.

Nowcasting comes with its own set of challenges that any modelling approach must be able to handle. First, the model should be able to handle time series. Second, it should have some mechanism for dealing with mixed-frequency data. This refers to the fact that the variables in the model, be it the target variable or input variables, will not necessarily be recorded in the same frequency, for instance estimating a yearly variable using monthly and quarterly variables. Third, it should be robust to the differing publication schedules of its input variables, often called "ragged-edges". The last challenge is the "curse of dimensionality", where there may be few observations relative to the number of input variables, complicating the estimation of many classical econometric and statistical models.

Several different methodologies address these challenges and have been used successfully for nowcasting applications. Some of the most common include the dynamic factor model (DFM) (Guichard and Rusticelli, 2011; Antolin-Diaz et al., 2020), mixed data sampling (MIDAS) (Kuzin et al., 2009; Marcellino and Schumacher, 2010), mixed data sampling vector autoregression (VAR) (Kuzin et al., 2009), and Bayesian vector autoregression (Cimadomo et al., 2020). Hopp (2021) examined an approach using long short-term memory artificial neural networks (LSTM).

No one approach is better than the others in all cases. Ideally, multiple would be tried in order to validate performance and increase the chances of obtaining a high-performing model. In practice, the methodology chosen will be influenced by other factors, such as which implementations are available in which programming languages, if any open-source options are available at all. It is primarily for this latter reason that the LSTM was chosen for this case study. In selecting a methodology, the *nowcasting_benchmark* open source repository is a good resource outlining the performance of all common nowcasting methodologies in nowcasting US GDP growth (Hopp, 2022). It additionally contains boilerplate code for each methodology which can be followed to perform one's own nowcasting exercise.

Having settled on the LSTM for the nowcasting model, the next steps are selecting which variables will go into the model and which hyperparameters will be used. If the number of input variables gathered in the data collection phase is small, the former may not be necessary. In order to accomplish this, a performance metric must first be determined to compare different models to each other. In a regression application such as this, mean absolute error (MAE) and/or root-mean-square error (RMSE) are suitable. For a given model with given input variables and hyperparameters, its accuracy needs to be assessed via the performance metric. In order to ensure that a model is generalizable and not overfit, it should be assessed on data it was not trained on. A general rule of thumb is to train on 80 per cent of the data and test on 20 per cent. In this case, models were trained on data ranging from 2001 to 2011 and validated on data from 2012-2014 (the validation period). They were then trained on data ranging from 2001-2014 and tested on data from 2015-2018 (the test period).

The logic behind the validation set is its use as a way of selecting variables and hyperparameters. Selecting the best performing models based on validation performance then finally assessing them on the test set ensures that the model is not being overfit on the test set in terms of its variables and hyperparameters.

A final factor to keep in mind in nowcasting is performance before all data are available. This is commonly the case in nowcasting, and especially so if the nowcast is to be monitored over a period of time. To account for this, model performance was recorded on synthetic data vintages, or the artificial introduction of missing values to simulate the data as it would have appeared at different points in time. The vintages corresponded to the month of the target period, i.e., if the target period was 2020, the data as it would have appeared in December 2020, six months after the target period, and ten months after the target period, when the latest publishing input variable would be released. Publication lags for generating the synthetic data vintages were gathered from empirical observation from April 2020 to October 2021.

Now with a specified process of training and testing a particular model, many models could be tested to determine the best performing one. Variable selection and a small degree of hyperparameter tuning were carried out in the same step for the case study. Because there were far too many potential input variables to test all permutations, input variables were randomly sampled, run with a small selection of hyperparameters, and their performance recorded. This process was repeated for hundreds of random input variable samples. The three best performing models of these runs were then assessed with a more expansive set of hyperparameters. Finally, the best performing of these was selected for the final model. There are other approaches to variable selection than the process described here, which will not lead to the absolute best performing input variable and hyperparameter combination possible from the data. This best performing combination is impractical to find due to computational and time constraints. However, this approach is sure to find a relatively well-performing model from the space of all possible input variable and hyperparameter combinations.

Results from the final selected model are presented in the next section.

## 4.3 Results

The variables selected for the final model are listed in table 1.

Table 1. Final selected variables

| Variable | Geography | Frequency | Source |
|---|---|---|---|
| construction index | France | monthly | OECD |
| consumer confidence index | Japan | monthly | OECD |
| goods volume transported by main ports | Netherlands | quarterly | Eurostat |
| manufacturing order books | Germany | monthly | OECD |
| merchandise exports | Singapore | monthly | Singapore DOS |
| merchandise exports | South Africa | monthly | OECD |
| real GDP forecast | OECD | quarterly | OECD |
| total energy consumed by transportation | USA | monthly | EIA |
| tourist arrivals | France | monthly | Eurostat |

Figure 1 shows the model's predictions for both the validation and test sets with full data compared with observed actuals. With these variables, full data is equivalent to about seven months after the period has ended, or July of the following year, when the variable with the longest lag is published. The blue line represents predictions on the validation set, so using a model trained with data from 2001 to only 2011. The red line represents predictions on the test set, with a model trained with data from 2001 to 2014. The model is remarkably accurate in the first two years of the validation set, 2012 and 2013, but struggles with the abnormally low observed 2014 level. The large drop in that year may have been due to abnormally warm winter weather in certain regions, reducing energy consumption, coupled with a reduction of coal use in China (Briggs, 2015; Scientific American, 2015). This suggests that inclusion of variables related to weather and fossil fuel use or prices could improve the performance of the model. Overestimation of 2014's value was common to all models, perhaps as the first year where carbon intensity of GDP levels began to decline at faster rates than previously observed in the ten years prior. However, the model was able to pick up on this faster declining trend between 2015 and 2017, as well as a relatively milder decline in 2018.

MAE and RMSE for the validation period were 0.011 and 0.024, respectively. In words, this means the model predicted year-over-year growth in the target variable that was 1.1 percentage points different from the actual over the validation period. The higher RMSE value shows how this particular performance metric punishes the larger error in 2014. MAE and RMSE for the test period were 0.006 and 0.007, respectively.

Figure 1. Indicator 9.4.1, carbon intensity of GDP

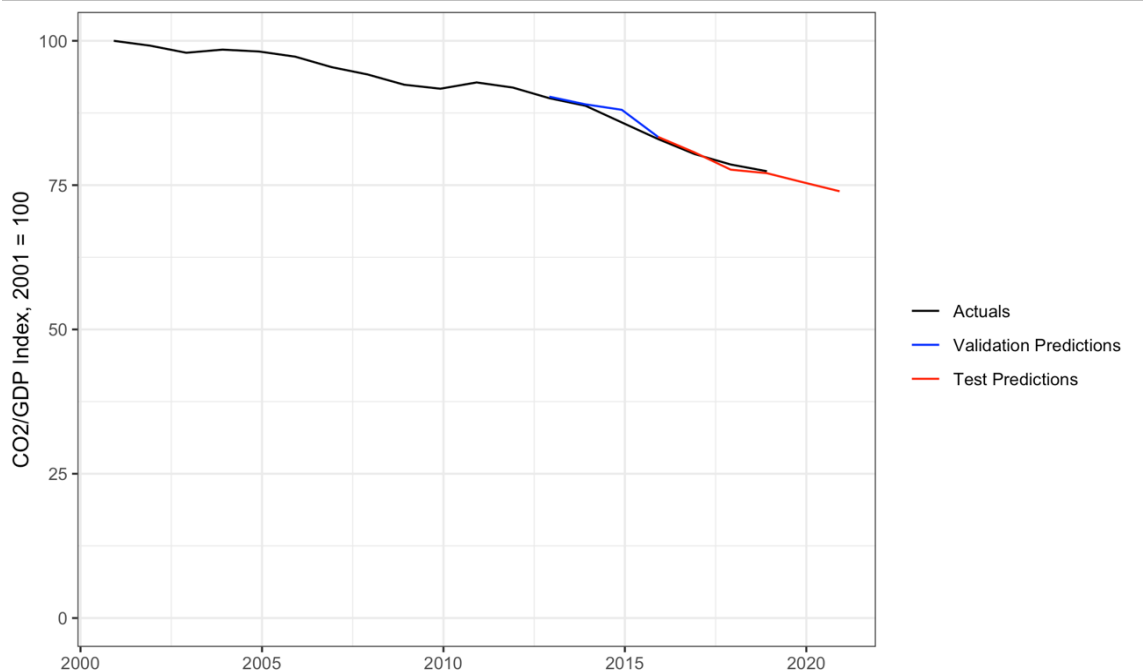

Figure 2 shows the development of 2015 and 2020's nowcasts over time, beginning in January of the target year, and ending in July of the following year. Predictions over time were gotten from running the model on synthetic data vintages based on the publication lags of the variables, or how the data would have looked at different points over the year. The y axis shows the nowcasted yearly growth rate, while the x axis shows the simulated date. Each point in the lines represents the nowcast for the yearly growth rate given the data that would have been available at that time. For 2015, we can observe a relatively monotonic development throughout the year, as the forecast was revised downwards as time went on. 2015 indeed registered the largest year over year decline in the target variable in the data; -3.3 per cent. In 2020, we can observe a development that changed directions over time, due to the volatile signals in the data owing to the COVID-19 pandemic. In the end, the model predicted reductions in line with the average for the period from 2014 to 2018, due to the fact that the pandemic affected both carbon emissions as well as economic activity. Figure 2 illustrates how nowcasting can be used to monitor the development of SDG indicators in real-time and gain insight to how various factors influence them.

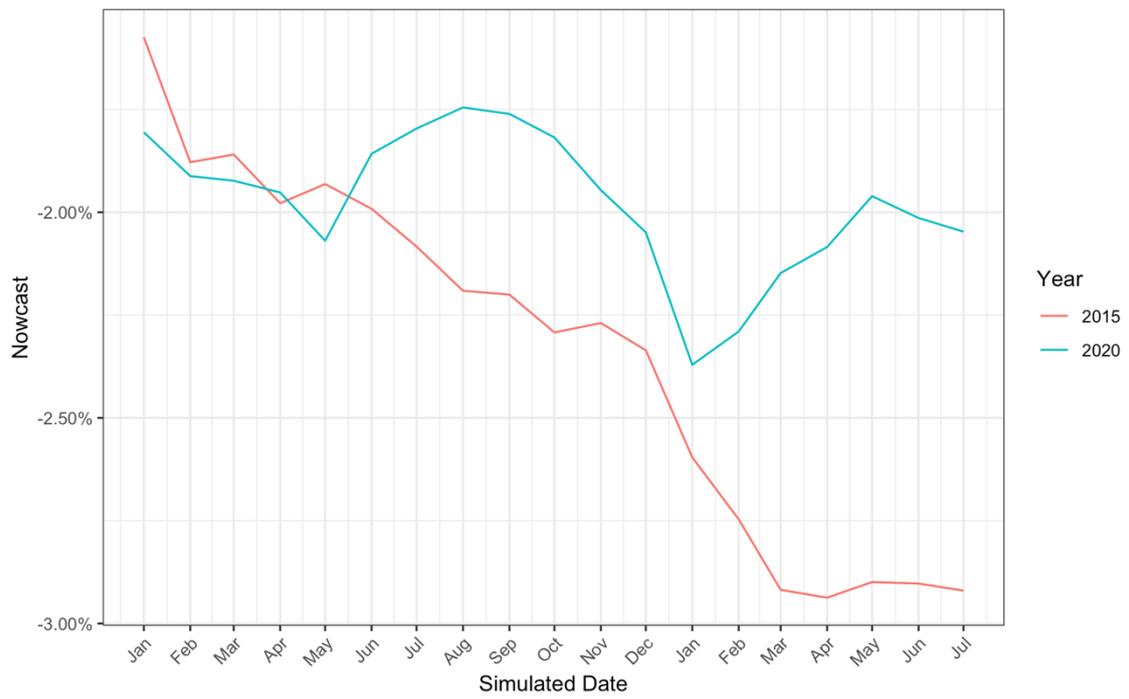

Figure 2. Development of nowcasts over time

# 5. Conclusion

Out of 362 Tier 1 SDG indicators and sub-indicators, we found the majority suitable for nowcasting purposes based on information available as of Autumn 2021. More specifically, 150 indicators were classified as "Highly likely" able to be nowcast, 87 were classified as "Likely", and 125 were classified as "Unlikely". See Appendix 1 for a visual overview of the nowcasting feasibility of sub-indicators by goal.

The case study conducted on indicator 9.4.1 illustrates the full process of nowcasting an SDG indicator. While the particular approach taken can vary considerably from that presented here, especially as it relates to the methodology employed, it can serve as a basis or starting point for those new to the practice.

While a large number of SDG indicators were considered highly feasible for nowcasting, about half of them were not. The paper only provides a first indication of the potential to investigate nowcasting feasibility. Nowcasting will not solve timeliness and data availability issues.
National efforts guided by the global statistical community and indicator custodian agencies are crucial to improving the availability of statistical data of sufficient quality, including time series length. For instance, work invested in backcasting time series to enable nowcasting and other efforts to increase the quality of SDG indicator data would also benefit nowcasting. Finally, it is the quality and availability of national statistical data that determines possibilities for nowcasting. One of the findings of this survey is that if and as policy makers require timelier data, more investment in official statistics, their quality and their comprehensiveness is needed.

As 2030 ticks nearer and the world looks to the challenges ahead, the SDG indicators will continue to be called upon for guidance. Nowcasting has the potential to increase the indicators' timeliness, and thus their usefulness. The survey and case study conducted in this paper hopefully contribute to crystallizing that potential by serving as resources for custodian agencies, national governments, or interested individuals in carrying out their own nowcasting exercises.

# Appendix

## Appendix 1. Nowcasting feasibility survey results overview by sub-indicator

*Note: Green highlight indicates "Highly likely", orange "Likely, and red "Unlikely*

| Goal 1 | Goal 2 | Goal 3 | Goal 4 | Goal 5 | Goal 6 | Goal 7 | Goal 8 | Goal 9 | Goal 10 | Goal 11 | Goal 12 | Goal 13 | Goal 14 | Goal 15 | Goal 16 | Goal 17 |
|---|---|---|---|---|---|---|---|---|---|---|---|---|---|---|---|---|
| 1.1.1 | 2.1.1 | 3.1.1 | 4.1.1 | 5.3.1 | 6.4.1 | 7.1.1 | 8.1.1 | 9.1.2 | 10.4.1 | 11.1.1 | 12.2.2 | 13.1.1 | 14.4.1 | 15.1.1 | 16.3.2 | 17.1.1 |
| 1.1.1 | 2.1.1 | 3.1.2 | 4.1.2 | 5.3.2 | 6.4.2 | 7.1.2 | 8.2.1 | 9.1.2 | 10.5.1 | 11.5.1 | 12.2.2 | 13.1.1 | 14.5.1 | 15.1.1 | 16.5.2 | 17.1.1 |
| 1.2.1 | 2.1.2 | 3.2.1 | 4.2.2 | 5.5.1 | 6.5.1 | 7.2.1 | 8.4.2 | 9.1.2 | 10.5.1 | 11.5.1 | 12.2.2 | 13.1.1 | 14.5.1 | 15.1.1 | 16.7.1 | 17.1.2 |
| 1.4.1 | 2.1.2 | 3.2.1 | 4.5.1 | 5.5.1 | 6.5.1 | 7.3.1 | 8.4.2 | 9.1.2 | 10.5.1 | 11.5.1 | 12.4.1 | 13.1.1 | 14.5.1 | 15.1.2 | 16.7.1 | 17.2.1 |
| 1.4.1 | 2.1.2 | 3.2.1 | 4.5.1 | 5.5.1 | 6.5.2 | 7.a.1 | 8.4.2 | 9.2.1 | 10.5.1 | 11.5.1 | 12.4.1 | 13.1.1 | 14.6.1 | 15.1.2 | 16.7.1 | 17.2.1 |
| 1.5.1 | 2.1.2 | 3.2.1 | 4.5.1 | 5.5.1 | 6.5.2 | 7.b.1 | 8.5.2 | 9.2.1 | 10.5.1 | 11.5.1 | 12.4.1 | 13.1.1 | 14.7.1 | 15.2.1 | 16.7.1 | 17.2.1 |
| 1.5.1 | 2.2.1 | 3.2.2 | 4.5.1 | 5.5.2 | 6.5.2 |  | 8.6.1 | 9.2.1 | 10.5.1 | 11.5.1 | 12.4.1 | 13.1.1 | 14.b.1 | 15.2.1 | 16.7.1 | 17.2.1 |
| 1.5.1 | 2.2.1 | 3.2.2 | 4.5.1 | 5.5.2 | 6.6.1 |  | 8.10.1 | 9.2.2 | 10.5.1 | 11.5.1 | 12.4.1 | 13.1.1 |  | 15.2.1 | 16.7.1 | 17.2.1 |
| 1.5.1 | 2.2.2 | 3.3.1 | 4.5.1 |  | 6.6.1 |  | 8.10.2 | 9.3.2 | 10.6.1 | 11.5.1 | 12.a.1 | 13.1.1 |  | 15.2.1 | 16.7.1 | 17.2.1 |
| 1.5.1 | 2.2.2 | 3.3.2 | 4.5.1 |  | 6.6.1 |  | 8.a.1 | 9.4.1 | 10.6.1 | 11.5.1 | 12.b.1 | 13.1.1 |  | 15.2.1 | 16.7.1 | 17.2.1 |
| 1.5.1 | 2.2.2 | 3.3.3 | 4.5.1 |  | 6.6.1 |  | 8.a.1 | 9.4.1 | 10.7.3 | 11.5.1 | 12.b.1 | 13.2.2 |  | 15.3.1 | 16.7.1 | 17.2.1 |
| 1.5.1 | 2.2.2 | 3.3.4 | 4.5.1 |  | 6.6.1 |  | 8.a.1 | 9.4.1 | 10.7.4 | 11.6.2 | 12.b.1 | 13.2.2 |  | 15.4.1 | 16.7.1 | 17.2.1 |
| 1.5.1 | 2.2.3 | 3.3.5 | 4.5.1 |  | 6.6.1 |  | 8.a.1 | 9.5.1 | 10.a.1 | 11.a.1 | 12.c.1 |  |  | 15.4.2 | 16.7.1 | 17.2.1 |
| 1.5.1 | 2.2.3 | 3.4.1 | 4.5.1 |  | 6.6.1 |  |  | 9.5.2 | 10.b.1 |  | 12.c.1 |  |  | 15.5.1 | 16.7.1 | 17.3.1 |
| 1.5.1 | 2.2.3 | 3.4.1 | 4.5.1 |  | 6.6.1 |  |  | 9.a.1 | 10.b.1 |  | 12.c.1 |  |  | 15.6.1 | 16.7.1 | 17.3.2 |
|  | 2.5.1 | 3.4.2 | 4.5.1 |  | 6.6.1 |  |  | 9.b.1 | 10.b.1 |  |  |  |  | 15.6.1 | 16.8.1 | 17.4.1 |
|  | 2.5.1 | 3.4.2 | 4.5.1 |  | 6.6.1 |  |  | 9.c.1 | 10.c.1 |  |  |  |  | 15.6.1 | 16.8.1 | 17.6.1 |
|  | 2.a.1 | 3.5.1 | 4.5.1 |  | 6.6.1 |  |  | 9.c.1 | 10.c.1 |  |  |  |  | 15.6.1 | 16.9.1 | 17.6.1 |
|  | 2.a.1 | 3.5.1 | 4.a.1 |  | 6.6.1 |  |  | 9.c.1 | 10.c.1 |  |  |  |  | 15.6.1 | 16.10.2 | 17.8.1 |
|  | 2.a.1 | 3.5.2 | 4.a.1 |  | 6.6.1 |  |  |  |  |  |  |  |  | 15.8.1 | 16.a.1 | 17.9.1 |
|  | 2.a.2 | 3.6.1 | 4.a.1 |  | 6.6.1 |  |  |  |  |  |  |  |  | 15.8.1 | 16.a.1 | 17.10.1 |
|  | 2.b.1 | 3.7.1 | 4.a.1 |  | 6.6.1 |  |  |  |  |  |  |  |  | 15.8.1 | 16.a.1 | 17.10.1 |
|  | 2.c.1 | 3.7.2 | 4.a.1 |  | 6.6.1 |  |  |  |  |  |  |  |  | 15.8.1 | 16.a.1 | 17.11.1 |
|  | 2.c.1 | 3.8.1 | 4.a.1 |  | 6.6.1 |  |  |  |  |  |  |  |  | 15.8.1 | 16.a.1 | 17.11.1 |
|  | 2.c.1 | 3.8.2 | 4.a.1 |  | 6.6.1 |  |  |  |  |  |  |  |  | 15.8.1 |  | 17.11.1 |
|  |  | 3.8.2 | 4.b.1 |  | 6.6.1 |  |  |  |  |  |  |  |  | 15.8.1 |  | 17.11.1 |
|  |  | 3.9.1 |  |  | 6.6.1 |  |  |  |  |  |  |  |  | 15.9.1 |  | 17.12.1 |
|  |  | 3.9.1 |  |  | 6.6.1 |  |  |  |  |  |  |  |  | 15.9.1 |  | 17.12.1 |
|  |  | 3.9.1 |  |  | 6.6.1 |  |  |  |  |  |  |  |  | 15.a.1 |  | 17.18.2 |
|  |  | 3.9.1 |  |  | 6.6.1 |  |  |  |  |  |  |  |  | 15.a.1 |  | 17.18.3 |
|  |  | 3.9.1 |  |  | 6.6.1 |  |  |  |  |  |  |  |  | 15.b.1 |  | 17.18.3 |
|  |  | 3.9.1 |  |  | 6.6.1 |  |  |  |  |  |  |  |  | 15.b.1 |  | 17.18.3 |
|  |  | 3.9.2 |  |  | 6.a.1 |  |  |  |  |  |  |  |  |  |  | 17.18.3 |
|  |  | 3.9.3 |  |  | 6.b.1 |  |  |  |  |  |  |  |  |  |  | 17.18.3 |
|  |  | 3.a.1 |  |  | 6.b.1 |  |  |  |  |  |  |  |  |  |  | 17.19.1 |
|  |  | 3.b.1 |  |  | 6.b.1 |  |  |  |  |  |  |  |  |  |  | 17.19.2 |
|  |  | 3.b.1 |  |  | 6.b.1 |  |  |  |  |  |  |  |  |  |  | 17.19.2 |
|  |  | 3.b.1 |  |  | 6.b.1 |  |  |  |  |  |  |  |  |  |  | 17.19.2 |
|  |  | 3.b.1 |  |  | 6.b.1 |  |  |  |  |  |  |  |  |  |  | 17.19.2 |
|  |  | 3.b.2 |  |  | 6.b.1 |  |  |  |  |  |  |  |  |  |  | 17.19.2 |
|  |  | 3.b.2 |  |  | 6.b.1 |  |  |  |  |  |  |  |  |  |  | 17.19.2 |
|  |  | 3.b.3 |  |  |  |  |  |  |  |  |  |  |  |  |  |  |
|  |  | 3.c.1 |  |  |  |  |  |  |  |  |  |  |  |  |  |  |
|  |  | 3.c.1 |  |  |  |  |  |  |  |  |  |  |  |  |  |  |
|  |  | 3.d.1 |  |  |  |  |  |  |  |  |  |  |  |  |  |  |

## Appendix 2. Complete nowcasting feasibility survey results

*Note: Only Tier 1 indicators included. All data refer to world aggregate unless otherwise specified. All data as of autumn 2021. More up-to-date or timelier data for individual indicators may be available directly from custodian indicators or elsewhere.*

| Indicator code | Name | Unit | Data availability | Publication lag | Explanatory variables | Nowcasting feasibility | Data source | Notes |
|---|---|---|---|---|---|---|---|---|
| 1.1.1 | Employed population below international poverty line, by sex and age | Percent | 2000-2017 | 2 years | Highly likely | Highly likely | Global SDG Indicator Database | |
| 1.1.1 | Proportion of population below international poverty line | Percent | 1981-2019 | 3 years | Highly likely | Highly likely | Global SDG Indicator Database | Length of time series for each country/region varies greatly. November/December data release. |
| 1.2.1 | Proportion of population living below the national poverty line, by sex and age | Percent | 1990-2015 | Data last released 2017 | Highly likely | Highly likely | Metadata 01-02-01 | |
| 1.4.1 | Proportion of population using basic drinking water services, by location | Percent | 1990-2020 | Varies by year | Highly likely | Highly likely | Metadata 01-04-01 | Data released every 3-5 years |
| 1.4.1 | Proportion of population using basic sanitation services, by location | Percent | 1990-2020 | Varies by year | Highly likely | Highly likely | Metadata 01-04-01 | Data released every 3-5 years |
| 1.5.1 | Number of deaths and missing persons attributed to disasters | Number | 2005-2020 | Likely 1 year | Highly likely | Likely | Global SDG Indicator Database | |
| 1.5.1 | Number of deaths and missing persons attributed to disasters per 100,000 population | Per 100,000 population | 2005-2020 | Likely 1 year | Highly likely | Likely | Global SDG Indicator Database | |
| 1.5.1 | Number of deaths due to disaster | Number | 2005-2020 | Likely 1 year | Highly likely | Likely | Global SDG Indicator Database | |
| 1.5.1 | Number of directly affected persons attributed to disasters per 100,000 population | Per 100,000 population | 2005-2020 | Likely 1 year | Highly likely | Likely | Global SDG Indicator Database | |

| Indicator code | Name | Unit | Data availability | Publication lag | Explanatory variables | Nowcasting feasibility | Data source | Notes |
|---|---|---|---|---|---|---|---|---|
| 1.5.1 | Number of injured or ill people attributed to disasters | Number | 2005-2020 | Likely 1 year | Highly likely | Likely | Global SDG Indicator Database | |
| 1.5.1 | Number of missing persons due to disaster | Number | 2005-2020 | Likely 1 year | Highly likely | Likely | Global SDG Indicator Database | |
| 1.5.1 | Number of people affected by disaster | Number | 2005-2020 | Likely 1 year | Highly likely | Likely | Global SDG Indicator Database | |
| 1.5.1 | Number of people whose damaged dwellings were attributed to disasters | Number | 2005-2020 | Likely 1 year | Highly likely | Likely | Global SDG Indicator Database | |
| 1.5.1 | Number of people whose destroyed dwellings were attributed to disasters | Number | 2005-2020 | Likely 1 year | Highly likely | Likely | Global SDG Indicator Database | |
| 1.5.1 | Number of people whose livelihoods were disrupted or destroyed, attributed to disasters | Number | 2005-2020 | Likely 1 year | Highly likely | Likely | Global SDG Indicator Database | |
| 2.1.1 | Number of undernourished people | Millions | 2001-2019 | Likely 2 years | Highly likely | Highly likely | Global SDG Indicator Database | Last data release July 2021 |
| 2.1.1 | Prevalence of undernourishment | Percent | 2001-2019 | Likely 2 years | Highly likely | Highly likely | Global SDG Indicator Database | Last data release July 2021 |
| 2.1.2 | Prevalence of moderate or severe food insecurity in the adult population | Percent | 2014-2019 | Likely 2 years | Highly likely | Unlikely | Global SDG Indicator Database | Last data release July 2021 |
| 2.1.2 | Prevalence of severe food insecurity in the adult population | Percent | 2014-2019 | Likely 2 years | Highly likely | Unlikely | Global SDG Indicator Database | Last data release July 2021 |
| 2.1.2 | Total population in moderate or severe food insecurity | Thousands | 2014-2019 | Likely 2 years | Highly likely | Unlikely | Global SDG Indicator Database | Last data release July 2021 |
| 2.1.2 | Total population in severe food insecurity | Thousands | 2014-2019 | Likely 2 years | Highly likely | Unlikely | Global SDG Indicator Database | Last data release July 2021 |
| 2.2.1 | Children moderately or severely stunted | Thousands | 2000-2020 | 1 year | Highly likely | Highly likely | Global SDG Indicator Database | May data release |

| Indicator code | Name | Unit | Data availability | Publication lag | Explanatory variables | Nowcasting feasibility | Data source | Notes |
|---|---|---|---|---|---|---|---|---|
| 2.2.1 | Proportion of children moderately or severely stunted | Percent | 2000-2020 | 1 year | Highly likely | Highly likely | Global SDG Indicator Database | May data release |
| 2.2.2 | Children moderately or severely overweight | Thousands | 2000-2020 | 1 year | Highly likely | Highly likely | Global SDG Indicator Database | May data release |
| 2.2.2 | Children moderately or severely wasted | Thousands | 2020 | 1 year | Highly likely | Unlikely | Global SDG Indicator Database | May data release |
| 2.2.2 | Proportion of children moderately or severely overweight | Percent | 2000-2020 | 1 year | Highly likely | Highly likely | Global SDG Indicator Database | May data release |
| 2.2.2 | Proportion of children moderately or severely wasted | Percent | 2020 | 1 year | Highly likely | Unlikely | Global SDG Indicator Database | May data release |
| 2.2.3 | Proportion of women aged 15-49 years with anaemia | Percent | 2000-2019 | Varies by year | Highly likely | Highly likely | Metadata 02-02-03 | Data released every 3-5 years |
| 2.2.3 | Proportion of women aged 15-49 years with anaemia, non-pregnant | Percent | 2000-2019 | Varies by year | Highly likely | Highly likely | Metadata 02-02-03 | Data released every 3-5 years |
| 2.2.3 | Proportion of women aged 15-49 years with anaemia | Percent | 2000-2019 | Varies by year | Highly likely | Highly likely | Metadata 02-02-03 | Data released every 3-5 years |
| 2.5.1 | Number of local breeds for which sufficient genetic resources are stored for reconstitution | Number | 2010-2021 | 1 year | Likely | Likely | Global SDG Indicator Database | No data available for most regions. 2021 data may be forecasted. Q1 data release. |
| 2.5.1 | Plant genetic resources accessions stored ex situ | Number | 1995, 2000, 2005, 2010, 2012, 2014, 2016-2020 | 1 year | Likely | Likely | Global SDG Indicator Database | No data available for most regions. Q1 data release. |
| 2.a.1 | Agriculture orientation index for government expenditures | Index | 2001-2019 | 2 years | Highly likely | Highly likely | Metadata 02-0a-01 | |
| 2.a.1 | Agriculture share of Government Expenditure | Percent | 2001-2019 | 2 years | Highly likely | Highly likely | Metadata 02-0a-01 | |
| 2.a.1 | Agriculture value added share of GDP | Percent | 2001-2019 | 2 years | Highly likely | Highly likely | Metadata 02-0a-01 | |

| Indicator code | Name | Unit | Data availability | Publication lag | Explanatory variables | Nowcasting feasibility | Data source | Notes |
|---|---|---|---|---|---|---|---|---|
| 2.a.2 | Total official flows (official development assistance plus other official flows) to the agriculture sector | Millions of constant 2019 United States dollars | 1973-2019 | 2 years | Likely | Likely | Metadata 02-0a-02 | December data release |
| 2.b.1 | Agricultural export subsidies | Millions of current United States dollars | 1995-2018 | Varies by WTO member | Highly likely | Highly likely | Metadata 02-0b-01, Global SDG Indicator Database | |
| 2.c.1 | Consumer Food Price Index | Index | January 2000-June 2021 | Varies by month | Highly likely | Highly likely | FAOSTAT | Monthly data released quarterly |
| 2.c.1 | Indicator of Food Price Anomalies (IFPA), by type of product | Index | 2015-2020 | Likely 1 year | Highly likely | Unlikely | Global SDG Indicator Database | Insufficient data for nowcasting purposes |
| 2.c.1 | Proportion of countries recording abnormally high or moderately high food prices, according to the Indicator of Food Price Anomalies | Percent | 2010-2019 | Likely 2 years | Highly likely | Unlikely | FAOSTAT | Insufficient data for nowcasting purposes |
| 3.1.1 | Maternal mortality ratio | Per 100,000 live births | 2000-2017 | Varies by data source | Highly likely | Highly likely | Metadata 03-01-01, Global SDG Indicator Database, WHO | |
| 3.1.2 | Proportion of births attended by skilled health personnel | Percent | 2000-2020 | Varies by household survey | Highly likely | Likely | Metadata 03-01-02, Global SDG Indicator Database, WHO | Length of time series for each country/region varies greatly |
| 3.2.1 | Infant deaths | Number | 1990-2019 | Likely 1 year | Highly likely | Highly likely | Metadata 03-02-01 | Last data release September 2020 |
| 3.2.1 | Infant mortality rate | Per 1,000 live births | 1990-2019 | Likely 1 year | Highly likely | Highly likely | Metadata 03-02-01 | Last data release September 2020 |
| 3.2.1 | Under-five deaths | Number | 1990-2019 | Likely 1 year | Highly likely | Highly likely | Metadata 03-02-01 | Last data release September 2020 |
| 3.2.1 | Under-five mortality rate, by sex | Per 1,000 live births | 1990-2019 | Likely 1 year | Highly likely | Highly likely | Metadata 03-02-01 | Last data release September 2020 |

| Indicator code | Name | Unit | Data availability | Publication lag | Explanatory variables | Nowcasting feasibility | Data source | Notes |
|---|---|---|---|---|---|---|---|---|
| 3.2.2 | Neonatal deaths | Number | 2000-2019 | Likely 1 year | Highly likely | Highly likely | Metadata 03-02-02, Global SDG Indicator Database | Last data release September 2020 |
| 3.2.2 | Neonatal mortality rate | Per 1,000 live births | 2000-2019 | Likely 1 year | Highly likely | Highly likely | Metadata 03-02-02, Global SDG Indicator Database | Last data release September 2020 |
| 3.3.1 | Number of new HIV infections by sex, age and key populations | Per 1,000 uninfected population | 2000-2019 | 2 years | Highly likely | Highly likely | Metadata 03-03-01, Global SDG Indicator Database | July data release |
| 3.3.2 | Tuberculosis incidence | Per 100,000 population | 2000-2019 | Likely 1-2 years | Highly likely | Highly likely | Metadata 03-03-02, Global SDG Indicator Database | October data release |
| 3.3.3 | Malaria incidence | Per 1,000 population | 2000-2019 | 2 years | Highly likely | Highly likely | Metadata 03-03-03, Global SDG Indicator Database | |
| 3.3.4 | Hepatitis B incidence | Per 100,000 population | 2015, 2018, 2020 | 2020 data released Q2 2021 | Highly likely | Unlikely | Metadata 03-03-04, Global SDG Indicator Database | Insufficient data for nowcasting purposes. Annual, Q2 data release. |
| 3.3.5 | Number of people requiring interventions against neglected tropical diseases | Number | 2010-2019 | Likely 2 years | Highly likely | Unlikely | Metadata 03-03-05, Global SDG Indicator Database | Insufficient data for nowcasting purposes. Last data release Q4 2020. |
| 3.4.1 | Mortality rate attributed to cardiovascular disease, cancer, diabetes or chronic respiratory disease | Probability | 2000, 2005, 2010, 2015, 2019 | Data last released 2020 | Highly likely | Unlikely | Metadata 03-04-01, Global SDG Indicator Database | Insufficient data for nowcasting purposes |
| 3.4.1 | Number of deaths attributed to non-communicable diseases, by type of disease and sex | Number | 2000, 2005, 2010, 2015, 2019 | Data last released 2020 | Highly likely | Unlikely | Metadata 03-04-01, Global SDG Indicator Database | Insufficient data for nowcasting purposes |
| 3.4.2 | Number of deaths attributed to suicide, by sex | Number | 2000, 2005, 2010, 2015, 2019 | Data last released 2020 | Highly likely | Unlikely | Metadata 03-04-02, Global SDG Indicator Database | Insufficient data for nowcasting purposes |
| 3.4.2 | Suicide mortality rate, by sex | Per 100,000 population | 2000, 2005, 2010, 2015, 2019 | Data last released 2020 | Highly likely | Unlikely | Metadata 03-04-02, Global SDG Indicator Database | Insufficient data for nowcasting purposes |

| Indicator code | Name | Unit | Data availability | Publication lag | Explanatory variables | Nowcasting feasibility | Data source | Notes |
|---|---|---|---|---|---|---|---|---|
| 3.5.1 | Alcohol use disorders, 12-month prevalence | Percent | 2016 | Data last released 2018 | Highly likely | Unlikely | WHO | Insufficient data for nowcasting purposes |
| 3.5.1 | Coverage of treatment interventions (pharmacological, psychosocial and rehabilitation and aftercare services) for substance use disorders | Percent | 2015-2017 | Data last released 2018 | Likely | Unlikely | WHO | Insufficient data for nowcasting purposes |
| 3.5.2 | Alcohol per capita consumption (aged 15 years and older) within a calendar year | Litres | 2000-2019 | Likely 2 years | Highly likely | Highly likely | WHO | |
| 3.6.1 | Death rate due to road traffic injuries | Per 100,000 population | 2000, 2005, 2010, 2015, 2019 | Data last released 2020 | Highly likely | Unlikely | Metadata 03-06-01, Global SDG Indicator Database | Insufficient data for nowcasting purposes |
| 3.7.1 | Proportion of women of reproductive age (aged 15–49 years) who have their need for family planning satisfied with modern methods | Percent | 2000-2021 | Likely 1-3 years | Highly likely | Highly likely | Metadata 03-07-01, Global SDG Indicator Database | Combination of data and model-based estimates. March data release. Length of time series for each country/region varies greatly |
| 3.7.2 | Adolescent birth rate (aged 10–14 years; aged 15–19 years) per 1,000 women in that age group | Per 1,000 women aged 15-19 years | 2000, 2005, 2010, 2015, 2020 | 1 year | Highly likely | Unlikely | Metadata 03-07-02, Global SDG Indicator Database | Q1 data release, five-year intervals |
| 3.8.1 | Coverage of essential health services | Index | 2000, 2005, 2010, 2015, 2017 | Data last released 2019 | Likely | Unlikely | Metadata 03-08-01, Global SDG Indicator Database | Insufficient data for nowcasting purposes |
| 3.8.2 | Proportion of population with large household expenditures on health (greater than 10%) as a share of total household expenditure or income | Percent | 2000-2018 | Varies by national statistical office | Highly likely | Likely | Metadata 03-08-02, Global SDG Indicator Database | Sporadic data availability at the country level. Length of time series for each country/region varies greatly |

39										UNCTAD Research Paper No. 82

| Indicator code | Name | Unit | Data availability | Publication lag | Explanatory variables | Nowcasting feasibility | Data source | Notes |
|---|---|---|---|---|---|---|---|---|
| 3.8.2 | Proportion of population with large household expenditures on health (greater than 25%) as a share of total household expenditure or income | Percent | 2000-2018 | Varies by national statistical office | Highly likely | Likely | Metadata 03-08-02, Global SDG Indicator Database | Sporadic data availability at the country level. Length of time series for each country/region varies greatly |
| 3.9.1 | Age-standardized mortality rate attributed to ambient air pollution | Per 100,000 population | 2016 | Data last released 2018 | Highly likely | Unlikely | WHO | Insufficient data for nowcasting purposes |
| 3.9.1 | Age-standardized mortality rate attributed to household air pollution | Per 100,000 population | 2016 | Data last released 2018 | Highly likely | Unlikely | WHO | Insufficient data for nowcasting purposes |
| 3.9.1 | Age-standardized mortality rate attributed to household and ambient air pollution | Per 100,000 population | 2016 | Data last released 2018 | Highly likely | Unlikely | WHO | Insufficient data for nowcasting purposes |
| 3.9.1 | Crude death rate attributed to ambient air pollution | Per 100,000 population | 2016 | Data last released 2018 | Highly likely | Unlikely | WHO | Insufficient data for nowcasting purposes |
| 3.9.1 | Crude death rate attributed to household air pollution | Per 100,000 population | 2016 | Data last released 2018 | Highly likely | Unlikely | WHO | Insufficient data for nowcasting purposes |
| 3.9.1 | Crude death rate attributed to household and ambient air pollution | Per 100,000 population | 2016 | Data last released 2018 | Highly likely | Unlikely | WHO | Insufficient data for nowcasting purposes |
| 3.9.2 | Mortality rate attributed to unsafe water, unsafe sanitation and lack of hygiene (exposure to unsafe Water, Sanitation and Hygiene for All (WASH) services) | Per 100,000 population | 2016 | Data last released 2017 | Highly likely | Unlikely | WHO | Insufficient data for nowcasting purposes |
| 3.9.3 | Mortality rate attributed to unintentional poisoning | Per 100,000 population | 2000, 2005, 2010, 2015, 2019 | Data last released 2020 | Highly likely | Unlikely | Metadata 03-09-03, Global SDG Indicator Database | Insufficient data for nowcasting purposes |
| 3.a.1 | Age-standardized prevalence of current tobacco use among persons aged 15 years and older | Percent | 2000, 2005, 2010, 2015-2018 | Varies by year | Highly likely | Unlikely | Metadata 03-0a-01, Global SDG Indicator Database | Biennial data release. Insufficient data for nowcasting purposes |

| Indicator code | Name | Unit | Data availability | Publication lag | Explanatory variables | Nowcasting feasibility | Data source | Notes |
|---|---|---|---|---|---|---|---|---|
| 3.b.1 | Proportion of the target population with access to 3 doses of diphtheria-tetanus-pertussis (DTP3) | Percent | 1980-2019 | 2 years | Highly likely | Highly likely | Metadata 03-0b-01, Global SDG Indicator Database | July data release |
| 3.b.1 | Proportion of the target population with access to affordable medicines and vaccines on a sustainable basis, human papillomavirus (HPV) | Percent | 1980-2019 | 2 years | Highly likely | Highly likely | Metadata 03-0b-01, Global SDG Indicator Database | July data release |
| 3.b.1 | Proportion of the target population with access to measles-containing-vaccine second-dose (MCV2) | Percent | 1980-2019 | 2 years | Highly likely | Highly likely | Metadata 03-0b-01, Global SDG Indicator Database | July data release |
| 3.b.1 | Proportion of the target population with access to pneumococcal conjugate 3rd dose (PCV3) | Percent | 1980-2019 | 2 years | Highly likely | Highly likely | Metadata 03-0b-01, Global SDG Indicator Database | July data release |
| 3.b.2 | Total official development assistance to medical research and basic heath sectors, gross disbursement, by recipient countries | Millions of constant 2019 United States dollars | 1973-2019 | 2 years | Likely | Likely | Metadata 03-0b-02, Global SDG Indicator Database | December data release |
| 3.b.2 | Total official development assistance to medical research and basic heath sectors, net disbursement, by recipient countries | Millions of constant 2019 United States dollars | 1973-2019 | 2 years | Likely | Likely | Metadata 03-0b-02, Global SDG Indicator Database | December data release |
| 3.b.3 | Proportion of health facilities that have a core set of relevant essential medicines available and affordable on a sustainable basis | Percent | 2004-2019 | Varies by country/region | Highly likely | Likely | Metadata 03-0b-03, Global SDG Indicator Database | Sporadic data availability at the country level. Length of time series for each country/region varies greatly |
| 3.c.1 | Health worker density, by type of occupation | Per 10,000 population | 2000-2018 | Likely 3 years | Highly likely | Highly likely | WHO | Last data release Q1 2021 |

| Indicator code | Name | Unit | Data availability | Publication lag | Explanatory variables | Nowcasting feasibility | Data source | Notes |
|---|---|---|---|---|---|---|---|---|
| 3.c.1 | Health worker distribution, by sex and type of occupation | Percent | 2000-2018 | Likely 3 years | Highly likely | Highly likely | WHO | Last data release Q1 2021 |
| 3.d.1 | International Health Regulations (IHR) capacity and health emergency preparedness | Percent | 2010-2020 | Varies by country/region | Highly likely | Likely | Metadata 03-0d-01, Global SDG Indicator Database | Length of time series for each country/region varies greatly |
| 4.1.1 | Proportion of children and young people (a) in grades 2/3; (b) at the end of primary; and (c) at the end of lower secondary achieving at least a minimum proficiency level in (i) reading and (ii) mathematics, by sex | Percent | 2000-2020 | Varies by country/region | Highly likely | Likely | Metadata 04-01-01, Global SDG Indicator Database | Sporadic data availability at the country level. Length of time series for each country/region varies greatly |
| 4.1.2 | Completion rate (primary education, lower secondary education, upper secondary education) | Percent | 2000-2019 | Varies by country/region | Highly likely | Likely | UIS, Global SDG Indicator Database | Sporadic data availability at the country level. Length of time series for each country/region varies greatly |
| 4.2.2 | Participation rate in organized learning (one year before the official primary entry age), by sex | Percent | 1998-2019 | Likely 2 years | Highly likely | Highly likely | UIS | |
| 4.5.1 | Adjusted gender parity index for achieving a minimum proficiency level in reading and mathematics | Ratio | 2000-2019 | Varies by country/region | Highly likely | Unlikely | Metadata 04-05-01, Global SDG Indicator Database | Sporadic data availability at the country level. Length of time series for each country/region varies greatly. Insufficient data for nowcasting purposes |
| 4.5.1 | Adjusted gender parity index for achieving at least a fixed level of proficiency in functional skills, by numeracy/literacy skills | Ratio | 2011, 2012, 2014, 2015, 2017 | Varies by country/region | Highly likely | Unlikely | sdg.org, Global SDG Indicator Database | Sporadic data availability at the country level. Insufficient data for nowcasting purposes |

| Indicator code | Name | Unit | Data availability | Publication lag | Explanatory variables | Nowcasting feasibility | Data source | Notes |
|---|---|---|---|---|---|---|---|---|
| 4.5.1 | Adjusted gender parity index for completion rate, by location, wealth quintile and education level | Index | 1996-2018 | Varies by country/region | Highly likely | Likely | sdg.org, Global SDG Indicator Database | Length of time series for each country/region varies greatly. Last data release 2020 |
| 4.5.1 | Adjusted gender parity index for participation rate in formal and non-formal education and training | Ratio | 2000-2018 | Varies by country/region | Highly likely | Likely | sdg.org, Global SDG Indicator Database | Length of time series for each country/region varies greatly. Last data release 2020 |
| 4.5.1 | Adjusted gender parity index for participation rate in organized learning (one year before the official primary entry age) | Ratio | 2000-2020 | Varies by country/region | Highly likely | Likely | sdg.org, Global SDG Indicator Database | Length of time series for each country/region varies greatly. Last data release 2021 |
| 4.5.1 | Adjusted gender parity index for the proportion of teachers with the minimum required qualifications, by education level | Ratio | 2000-2020 | Varies by country/region | Highly likely | Likely | sdg.org, Global SDG Indicator Database | Length of time series for each country/region varies greatly. Last data release 2021 |
| 4.5.1 | Adjusted immigration status parity index for achieving a minimum proficiency level in reading and mathematics | Ratio | 2000-2018 | Varies by country/region | Highly likely | Unlikely | sdg.org, Global SDG Indicator Database | Sporadic data availability at the country level. Insufficient data for nowcasting purposes |
| 4.5.1 | Adjusted immigration status parity index for achieving at least a fixed level of proficiency in functional skills, by numeracy/literacy skills | Ratio | 2000-2018 | Varies by country/region | Highly likely | Unlikely | sdg.org, Global SDG Indicator Database | Sporadic data availability at the country level. Insufficient data for nowcasting purposes |
| 4.5.1 | Adjusted language test parity index for achieving a minimum proficiency level in reading and mathematics | Ratio | 2000-2018 | Varies by country/region | Highly likely | Unlikely | sdg.org, Global SDG Indicator Database | Sporadic data availability at the country level. Insufficient data for nowcasting purposes |
| 4.5.1 | Adjusted location parity index for completion rate, by sex, wealth quintile and education level | Index | 2000-2018 | Varies by country/region | Highly likely | Unlikely | sdg.org, Global SDG Indicator Database | Sporadic data availability at the country level. Insufficient data for nowcasting purposes |

| Indicator code | Name | Unit | Data availability | Publication lag | Explanatory variables | Nowcasting feasibility | Data source | Notes |
|---|---|---|---|---|---|---|---|---|
| 4.5.1 | Adjusted low to high socio-economic parity index for achieving a minimum proficiency level in reading and mathematics | Ratio | 2000-2018 | Varies by country/region | Highly likely | Unlikely | sdg.org, Global SDG Indicator Database | Length of time series for each country/region varies greatly. Insufficient data for nowcasting purposes |
| 4.5.1 | Adjusted low to high socio-economic parity status index for achieving at least a fixed level of proficiency in functional skills, by numeracy/literacy skills | Ratio | 2012, 2014, 2015, 2017 | Varies by country/region | Highly likely | Unlikely | sdg.org, Global SDG Indicator Database | Sporadic data availability at the country level. Insufficient data for nowcasting purposes |
| 4.5.1 | Adjusted rural to urban parity index for achieving a minimum proficiency level in reading and mathematics | Ratio | 2000-2018 | Varies by country/region | Highly likely | Unlikely | sdg.org, Global SDG Indicator Database | Sporadic data availability at the country level. Insufficient data for nowcasting purposes |
| 4.5.1 | Adjusted wealth parity index for completion rate, by sex, location and education level | Index | 2000-2019 | Varies by country/region | Highly likely | Likely | sdg.org, Global SDG Indicator Database | Length of time series for each country/region varies greatly. Last data release 2020 |
| 4.5.1 | Gender parity index for youth/adults with information and communications technology (ICT) skills, by type of skill | Ratio | 2010-2020 | Varies by country/region | Highly likely | Likely | Metadata 04-05-01, Global SDG Indicator Database | Length of time series for each country/region varies greatly |
| 4.a.1 | Proportion of schools with access to adapted infrastructure and materials for students with disabilities, by education level | Percent | 2015-2020 | Varies by country/region | Highly likely | Unlikely | Metadata 04-0a-01, Global SDG Indicator Database | Sporadic data availability at the country level. Insufficient data for nowcasting purposes. |
| 4.a.1 | Proportion of schools with access to basic drinking water, by education level | Percent | 2000-2020 | Varies by country/region | Highly likely | Likely | Metadata 04-0a-01, Global SDG Indicator Database | Length of time series for each country/region varies greatly |
| 4.a.1 | Proportion of schools with access to the internet for pedagogical purposes, by education level | Percent | 2000-2019 | Varies by country/region | Highly likely | Likely | Metadata 04-0a-01, Global SDG Indicator Database | Length of time series for each country/region varies greatly |

| Indicator code | Name | Unit | Data availability | Publication lag | Explanatory variables | Nowcasting feasibility | Data source | Notes |
|---|---|---|---|---|---|---|---|---|
| 4.a.1 | Proportion of schools with access to access to single-sex basic sanitation, by education level | Percent | 2000-2020 | Varies by country/region | Highly likely | Likely | Metadata 04-0a-01, Global SDG Indicator Database | Length of time series for each country/region varies greatly |
| 4.a.1 | Proportion of schools with access to computers for pedagogical purposes, by education level | Percent | 2000-2020 | Varies by country/region | Highly likely | Likely | Metadata 04-0a-01, Global SDG Indicator Database | Length of time series for each country/region varies greatly |
| 4.a.1 | Proportion of schools with access to electricity, by education level | Percent | 2000-2020 | Varies by country/region | Highly likely | Likely | Metadata 04-0a-01, Global SDG Indicator Database | Length of time series for each country/region varies greatly |
| 4.a.1 | Proportion of schools with basic handwashing facilities, by education level | Percent | 2000-2020 | Varies by country/region | Highly likely | Likely | Metadata 04-0a-01, Global SDG Indicator Database | Length of time series for each country/region varies greatly |
| 4.b.1 | Volume of official development assistance flows for scholarships by sector and type of study | Millions of constant 2019 United States dollars | 1998-2019 | Varies by country/region | Likely | Likely | UIS | |
| 5.3.1 | Proportion of women aged 20–24 years who were married or in a union before age 15 and before age 18 | Percent | 2000-2020 | Varies by data source | Highly likely | Unlikely | Metadata 05-03-01 | Sporadic data availability at the country level. Insufficient data for nowcasting purposes |
| 5.3.2 | Proportion of girls and women aged 15–49 years who have undergone female genital mutilation/cutting, by age | Percent | 2000-2020 | Varies by data source | Highly likely | Likely | Metadata 05-03-02 | Sporadic data availability at the country level |
| 5.5.1 | Current number of seats in national parliaments | Number | 1997-2021 | Likely 1 month | Unlikely | Unlikely | IPU rankings | Monthly data release |
| 5.5.1 | Number of seats held by women in national parliaments | Number | 1997-2021 | Likely 1 month | Likely | Likely | IPU rankings | Monthly data release |

| Indicator code | Name | Unit | Data availability | Publication lag | Explanatory variables | Nowcasting feasibility | Data source | Notes |
|---|---|---|---|---|---|---|---|---|
| 5.5.1 | Proportion of elected seats held by women in deliberative bodies of local government | Percent | 1980, 1990, 1995, 2000-2020 | 1 year | Likely | Likely | UNECE | Q2 data release |
| 5.5.1 | Proportion of seats held by women in national parliaments | Percent | 1997-2021 | Likely 1 month | Likely | Likely | IPU rankings | Monthly data release |
| 5.5.2 | Proportion of women in managerial positions | Percent | 1991-2020 | Likely 1 year | Highly likely | Highly likely | ILOSTAT | |
| 5.5.2 | Proportion of women in senior and middle management positions | Percent | 2000-2020 | Likely 1 year | Highly likely | Highly likely | ILOSTAT | |
| 6.4.1 | Change in water-use efficiency over time | United States dollars per cubic meter | 1961-2018 | Likely 3 years | Highly likely | Highly likely | Metadata-06-04-01 | Length of time series for each country/region varies greatly |
| 6.4.2 | Level of water stress: freshwater withdrawal as a proportion of available freshwater resources | Percent | 1961-2018 | Likely 3 years | Highly likely | Highly likely | Metadata-06-04-02 | Length of time series for each country/region varies greatly. February data release |
| 6.5.1 | Degree of integrated water resources management implementation | Percent | 2017, 2020 | Varies by year | Likely | Unlikely | Metadata 06-05-01, Global SDG Indicator Database | Data released every 3-5 years |
| 6.5.1 | Proportion of countries by IWRM implementation category | Percent | 2017, 2020 | Varies by year | Likely | Unlikely | Metadata 06-05-01, Global SDG Indicator Database | Data released every 3-5 years |
| 6.5.2 | Proportion of transboundary aquifers with an operational arrangement for water cooperation | Percent | 2017, 2020 | 1 year | Likely | Unlikely | Metadata 06-05-02, Global SDG Indicator Database | Data released every 3 years |
| 6.5.2 | Proportion of transboundary basins (river and lake basins and aquifers) with an operational arrangement for water cooperation | Percent | 2017, 2020 | 1 year | Likely | Unlikely | Metadata 06-05-02, Global SDG Indicator Database | Data released every 3 years |

| Indicator code | Name | Unit | Data availability | Publication lag | Explanatory variables | Nowcasting feasibility | Data source | Notes |
|---|---|---|---|---|---|---|---|---|
| 6.5.2 | Proportion of transboundary river and lake basins with an operational arrangement for water cooperation | Percent | 2017, 2020 | 1 year | Likely | Unlikely | Metadata 06-05-02, Global SDG Indicator Database | Data released every 3 years |
| 6.6.1 | Extent of human made wetlands | Square kilometres | 1984-2019 | Likely 2 years | Highly likely | Highly likely | UNEP, Metadata 06-06-01 | |
| 6.6.1 | Extent of inland wetlands | Square kilometres | 1984-2019 | Likely 2 years | Highly likely | Highly likely | UNEP, Metadata 06-06-01 | |
| 6.6.1 | Lake water quality trophic state | Percent | 1984-2019 | Likely 2 years | Highly likely | Highly likely | UNEP, Metadata 06-06-01 | |
| 6.6.1 | Lake water quality turbidity | Percent | 1984-2019 | Likely 2 years | Highly likely | Highly likely | UNEP, Metadata 06-06-01 | |
| 6.6.1 | Lakes and rivers permanent water area | Percent | 1984-2019 | Likely 2 years | Highly likely | Highly likely | UNEP, Metadata 06-06-01 | |
| 6.6.1 | Lakes and rivers permanent water area | Square kilometres | 1984-2019 | Likely 2 years | Highly likely | Highly likely | UNEP, Metadata 06-06-01 | |
| 6.6.1 | Lakes and rivers permanent water area change | Percent | 1984-2019 | Likely 2 years | Highly likely | Highly likely | UNEP, Metadata 06-06-01 | |
| 6.6.1 | Lakes and rivers seasonal water area | Percent | 1984-2019 | Likely 2 years | Highly likely | Highly likely | UNEP, Metadata 06-06-01 | |
| 6.6.1 | Lakes and rivers seasonal water area | Square kilometres | 1984-2019 | Likely 2 years | Highly likely | Highly likely | UNEP, Metadata 06-06-01 | |
| 6.6.1 | Lakes and rivers seasonal water area change | Percent | 1984-2019 | Likely 2 years | Highly likely | Highly likely | UNEP, Metadata 06-06-01 | |
| 6.6.1 | Mangrove area | Square kilometres | 1984-2019 | Likely 2 years | Highly likely | Highly likely | UNEP, Metadata 06-06-01 | |

| Indicator code | Name | Unit | Data availability | Publication lag | Explanatory variables | Nowcasting feasibility | Data source | Notes |
|---|---|---|---|---|---|---|---|---|
| 6.6.1 | Mangrove area baseline | Square kilometres | 1984-2019 | Likely 2 years | Highly likely | Highly likely | UNEP, Metadata 06-06-01 | |
| 6.6.1 | Mangrove area gain | Percent | 1984-2019 | Likely 2 years | Highly likely | Highly likely | UNEP, Metadata 06-06-01 | |
| 6.6.1 | Mangrove area gain | Square kilometres | 1984-2019 | Likely 2 years | Highly likely | Highly likely | UNEP, Metadata 06-06-01 | |
| 6.6.1 | Mangrove area loss | Percent | 1984-2019 | Likely 2 years | Highly likely | Highly likely | UNEP, Metadata 06-06-01 | |
| 6.6.1 | Mangrove area loss | Square kilometres | 1984-2019 | Likely 2 years | Highly likely | Highly likely | UNEP, Metadata 06-06-01 | |
| 6.6.1 | Mangrove total area change | Percent | 1984-2019 | Likely 2 years | Highly likely | Highly likely | UNEP, Metadata 06-06-01 | |
| 6.6.1 | Nationally derived quantity of groundwater | Millions of cubic metres per annum | 1984-2019 | Likely 2 years | Highly likely | Highly likely | UNEP, Metadata 06-06-01 | |
| 6.6.1 | Nationally derived quantity of rivers | Millions of cubic metres per annum | 1984-2019 | Likely 2 years | Highly likely | Highly likely | UNEP, Metadata 06-06-01 | |
| 6.6.1 | Reservoir maximum water area | Percent | 1984-2019 | Likely 2 years | Highly likely | Highly likely | UNEP, Metadata 06-06-01 | |
| 6.6.1 | Reservoir maximum water area | Square kilometres | 1984-2019 | Likely 2 years | Highly likely | Highly likely | UNEP, Metadata 06-06-01 | |
| 6.6.1 | Reservoir minimum water area | Percent | 1984-2019 | Likely 2 years | Highly likely | Highly likely | UNEP, Metadata 06-06-01 | |
| 6.6.1 | Reservoir minimum water area | Square kilometres | 1984-2019 | Likely 2 years | Highly likely | Highly likely | UNEP, Metadata 06-06-01 | |

| Indicator code | Name | Unit | Data availability | Publication lag | Explanatory variables | Nowcasting feasibility | Data source | Notes |
|---|---|---|---|---|---|---|---|---|
| 6.6.1 | Wetlands area | Percent | 1984-2019 | Likely 2 years | Highly likely | Highly likely | UNEP, Metadata 06-06-01 | |
| 6.6.1 | Wetlands area | Square kilometres | 1984-2019 | Likely 2 years | Highly likely | Highly likely | UNEP, Metadata 06-06-01 | |
| 6.a.1 | Amount of water- and sanitation-related official development assistance that is part of a government-coordinated spending plan | Millions of constant 2019 United States dollars | 2000-2019 | Likely 2 years | Likely | Likely | WHO, OECD | Last data release 2021 |
| 6.b.1 | Countries with procedures in law or policy for participation by service users/communities in planning program in rural drinking-water supply, by level of definition in procedures | 10 = Clearly defined; 5 = Not clearly defined ; 0 = NA | 2010-2019 | Likely 2 years | Likely | Unlikely | WHO | Length of time series for each country/region varies greatly. Insufficient data for nowcasting purposes |
| 6.b.1 | Countries with procedures in law or policy for participation by service users/communities in planning program in water resources planning and management, by level of definition in procedures | 10 = Clearly defined; 5 = Not clearly defined ; 0 = NA | 2017, 2019 | Likely 2 years | Likely | Unlikely | WHO | Insufficient data for nowcasting purposes |
| 6.b.1 | Countries with users/communities participating in planning programs in rural drinking-water supply, by level of participation | 3 = High; 2 = Moderate; 1 = Low; 0 = NA | 2010-2019 | Varies by country/region | Likely | Unlikely | WHO | Length of time series for each country/region varies greatly. Insufficient data for nowcasting purposes |
| 6.b.1 | Countries with users/communities participating in planning programs in water resources planning and management, by level of participation | 3 = High; 2 = Moderate; 1 = Low; 0 = NA | 2017, 2019 | Likely 2 years | Likely | Unlikely | WHO | Insufficient data for nowcasting purposes |

| Indicator code | Name | Unit | Data availability | Publication lag | Explanatory variables | Nowcasting feasibility | Data source | Notes |
|---|---|---|---|---|---|---|---|---|
| 6.b.1 | Proportion of countries with clearly defined procedures in law or policy for participation by service users/communities in planning program in rural drinking-water supply | Percent | 2012, 2014, 2017, 2019 | Likely 2 years | Likely | Unlikely | WHO | Insufficient data for nowcasting purposes |
| 6.b.1 | Proportion of countries with clearly defined procedures in law or policy for participation by service users/communities in planning program in water resources planning and management | Percent | 2017, 2019 | Likely 2 years | Likely | Unlikely | WHO | Insufficient data for nowcasting purposes |
| 6.b.1 | Proportion of countries with high level of users/communities participating in planning programs in rural drinking-water supply | Percent | 2014, 2017, 2019 | Likely 2 years | Likely | Unlikely | WHO | Insufficient data for nowcasting purposes |
| 6.b.1 | Proportion of countries with high level of users/communities participating in planning programs in water resources planning and management | Percent | 2017, 2019 | Likely 2 years | Likely | Unlikely | WHO | Insufficient data for nowcasting purposes |
| 7.1.1 | Proportion of population with access to electricity | Percent | 1990-2019 | 2 years | Highly likely | Highly likely | Metadata 07-01-01, Global SDG Indicators Database | |
| 7.1.2 | Proportion of population with primary reliance on clean fuels and technology | Percent | 1960-2019 | 2 years | Highly likely | Highly likely | Metadata 07-01-02, Global SDG Indicators Database | |
| 7.2.1 | Renewable energy share in the total final energy consumption | Percent | 2000-2018 | 2 years | Highly likely | Highly likely | Metadata 07-02-01, Global SDG Indicators Database | |

| Indicator code | Name | Unit | Data availability | Publication lag | Explanatory variables | Nowcasting feasibility | Data source | Notes |
|---|---|---|---|---|---|---|---|---|
| 7.3.1 | Energy intensity measured in terms of primary energy and GDP | Megajoules per constant 2017 purchasing power parity GDP | 2000-2018 | 2 years | Highly likely | Highly likely | Metadata 07-03-01, Global SDG Indicators Database | |
| 7.a.1 | International financial flows to developing countries in support of clean energy research and development and renewable energy production, including in hybrid systems | Millions of constant United States dollars | 2000-2018 | 2 years | Likely | Likely | Metadata 07-0a-01, Global SDG Indicators Database | |
| 7.b.1 | Installed renewable energy-generating capacity in developing countries | Watts per capita | 2000-2019 | 1 year | Highly likely | Highly likely | Metadata 07-0b-01, Global SDG Indicators Database | |
| 8.1.1 | Annual growth rate of real GDP per capita | Percent | 1970-2019 | 1 year | Highly likely | Highly likely | Metadata 08-01-01, Global SDG Indicators Database | |
| 8.2.1 | Annual growth rate of real GDP per employed person | Percent | 1991-2019 | 1 year | Highly likely | Highly likely | ILOSTAT | |
| 8.4.2 | Domestic material consumption per capita, by type of raw material | Tonnes | 1970-2017 | Last release 2017 | Highly likely | Highly likely | Metadata 08-04-02, Global SDG Indicators Database | |
| 8.4.2 | Domestic material consumption per unit of GDP, by type of raw material | Kilograms per constant 2010 United States dollars | 1970-2017 | Last release 2017 | Highly likely | Highly likely | Metadata 08-04-02, Global SDG Indicators Database | |
| 8.4.2 | Domestic material consumption, by type of raw material | Tonnes | 1970-2017 | Last release 2017 | Highly likely | Highly likely | Metadata 08-04-02, Global SDG Indicators Database | |

| Indicator code | Name | Unit | Data availability | Publication lag | Explanatory variables | Nowcasting feasibility | Data source | Notes |
|---|---|---|---|---|---|---|---|---|
| 8.5.2 | Unemployment rate, by sex, age and persons with disabilities | Percent | 2000-2019 (by sex and age), 2005-2019 (by disability) | Varies by national statistical office | Highly likely | Highly likely | Metadata 08-05-02, Global SDG Indicators Database | |
| 8.6.1 | Proportion of youth (aged 15–24 years) not in education, employment or training | Percent | 2004-2020 | Varies by national statistical office | Highly likely | Highly likely | ILOSTAT | Length of time series for each country/region varies greatly |
| 8.10.1 | (a) Number of commercial bank branches per 100,000 adults and (b) number of automated teller machines (ATMs) per 100,000 adults | Per 100,000 adults | 2004-2019 | Varies by financial regulator | Likely | Likely | Metadata 08-10-01, Global SDG Indicators Database | Length of time series for each country/region varies greatly |
| 8.10.2 | Proportion of adults (15 years and older) with an account at a bank or other financial institution or with a mobile-money-service provider | Percent | 2011, 2014, 2017 | 3 years | Highly likely | Unlikely | Metadata 08-10-02, Global SDG Indicators Database, World Bank | Data released every three years. Insufficient data for nowcasting purposes |
| 8.a.1 | Total official flows (commitments) for Aid for Trade, by donor countries | Millions of constant 2019 United States dollars | 2005-2019 | 1 year | Likely | Likely | Metadata 08-0a-01, Global SDG Indicators Database | |
| 8.a.1 | Total official flows (commitments) for Aid for Trade, by recipient countries | Millions of constant 2019 United States dollars | 2005-2019 | 1 year | Likely | Likely | Metadata 08-0a-01, Global SDG Indicators Database | |
| 8.a.1 | Total official flows (disbursement) for Aid for Trade, by donor countries | Millions of constant 2019 United States dollars | 2005-2019 | 1 year | Likely | Likely | Metadata 08-0a-01, Global SDG Indicators Database | |

| Indicator code | Name | Unit | Data availability | Publication lag | Explanatory variables | Nowcasting feasibility | Data source | Notes |
|---|---|---|---|---|---|---|---|---|
| 8.a.1 | Total official flows (disbursement) for Aid for Trade, by recipient countries | Millions of constant 2019 United States dollars | 2005-2019 | 1 year | Likely | Likely | Metadata 08-0a-01, Global SDG Indicators Database | |
| 9.1.2 | Container port traffic, maritime transport | Twenty-foot equivalent units | 2010-2019 | 1/2 years | Highly likely | Unlikely | UNCTAD | Insufficient data for nowcasting purposes |
| 9.1.2 | Freight loaded and unloaded, maritime transport | Metric tons | 2015-2019 | 1/2 years | Highly likely | Unlikely | Metadata 09-01-02, Global SDG Indicators Database | Insufficient data for nowcasting purposes |
| 9.1.2 | Freight volume, by mode of transport | Tonne kilometres | 1970-2019 | 1/2 years | Highly likely | Highly likely | OECD | |
| 9.1.2 | Passenger volume, by mode of transport | Passenger kilometres | 1998-2019 | 1/2 years | Highly likely | Highly likely | OECD | |
| 9.2.1 | Manufacturing value added (constant 2015 United States dollars) as a proportion of GDP | Percent | 2000-2020 | 1 year | Highly likely | Highly likely | Metadata 09-02-01, Global SDG Indicators Database | Data are already nowcasted by UNIDO to enhance a timely analysis of manufacturing trends |
| 9.2.1 | Manufacturing value added (current United States dollars) as a proportion of GDP | Percent | 2000-2020 | 1 year | Highly likely | Highly likely | Metadata 09-02-01, Global SDG Indicators Database | Data are already nowcasted by UNIDO to enhance a timely analysis of manufacturing trends |
| 9.2.1 | Manufacturing value added per capita | Constant 2015 United States dollars | 2000-2020 | 1 year | Highly likely | Highly likely | Metadata 09-02-01, Global SDG Indicators Database | Data are already nowcasted by UNIDO to enhance a timely analysis of manufacturing trends |
| 9.2.2 | Manufacturing employment as a proportion of total employment | Percent | 2000-2019 | Varies by national statistical office | Highly likely | Highly likely | Metadata 09-02-02, Global SDG Indicators Database | |

| Indicator code | Name | Unit | Data availability | Publication lag | Explanatory variables | Nowcasting feasibility | Data source | Notes |
|---|---|---|---|---|---|---|---|---|
| 9.3.2 | Proportion of small-scale industries with a loan or line of credit | Percent | 2006-2020 | Varies by country/region | Highly likely | Highly likely | Metadata 09-03-02, Global SDG Indicators Database | Sporadic data availability at the country level |
| 9.4.1 | Carbon dioxide emissions from fuel combustion | Millions of tonnes | 2000-2020 | 2 years | Highly likely | Highly likely | Metadata 09-04-01, Global SDG Indicators Database | Sporadic data availability at the country level |
| 9.4.1 | Carbon dioxide emissions per unit of GDP | Kilogrammes of $CO_2$ per constant 2017 United States dollars | 2000-2020 | 2 years | Highly likely | Highly likely | Metadata 09-04-01, Global SDG Indicators Database | Sporadic data availability at the country level |
| 9.4.1 | Carbon dioxide emissions per unit of manufacturing value added | Kilogrammes of $CO_2$ per constant 2015 United States dollars | 2000-2020 | 2 years | Highly likely | Highly likely | Metadata 09-04-01, Global SDG Indicators Database | Sporadic data availability at the country level |
| 9.5.1 | Research and development expenditure as a proportion of GDP | Percent | 1981-2019 | 2 years | Highly likely | Highly likely | Metadata 09-05-01, Global SDG Indicators Database | |
| 9.5.2 | Researchers (in full-time equivalent) per million inhabitants | Per 1,000,000 population | 1996-2019 | 2 years | Highly likely | Highly likely | UNESCO | |
| 9.a.1 | Total official international support (official development assistance plus other official flows) to infrastructure | Millions of constant 2019 United States dollars | 2002-2019 | 1 year | Likely | Likely | Metadata 09-0a-01, Global SDG Indicators Database | December data release |
| 9.b.1 | Proportion of medium and high-tech industry value added in total value added | Percent | 2000-2018 | 1 year | Highly likely | Highly likely | Metadata 09-0b-01, Global SDG Indicators Database | |
| 9.c.1 | Proportion of population covered by at least a 2G mobile network | Percent | 2000-2019 | 1 year | Highly likely | Highly likely | Metadata 09-0c-01, Global SDG Indicators Database | |

| Indicator code | Name | Unit | Data availability | Publication lag | Explanatory variables | Nowcasting feasibility | Data source | Notes |
|---|---|---|---|---|---|---|---|---|
| 9.c.1 | Proportion of population covered by at least a 3G mobile network | Percent | 2000-2019 | 1 year | Highly likely | Highly likely | Metadata 09-0c-01, Global SDG Indicators Database | |
| 9.c.1 | Proportion of population covered by at least a 4G mobile network | Percent | 2000-2019 | 1 year | Highly likely | Highly likely | Metadata 09-0c-01, Global SDG Indicators Database | |
| 10.4.1 | Labour share of GDP | Percent | 2004-2017 | Varies by year | Highly likely | Highly likely | Metadata 10-04-01, Global SDG Indicators Database | Biennial data release |
| 10.5.1 | Liquid assets to short term liabilities | Percent | 2005-2020 | Varies by country/region | Highly likely | Highly likely | Metadata 10-05-01, Global SDG Indicators Database | |
| 10.5.1 | Net open position in foreign exchange to capital | Percent | 2005-2020 | Varies by country/region | Highly likely | Highly likely | Metadata 10-05-01, Global SDG Indicators Database | |
| 10.5.1 | Non-performing loans net of provisions to capital | Percent | 2005-2020 | Varies by country/region | Highly likely | Highly likely | Metadata 10-05-01, Global SDG Indicators Database | |
| 10.5.1 | Non-performing loans to total gross loans | Percent | 2005-2020 | Varies by country/region | Highly likely | Highly likely | Metadata 10-05-01, Global SDG Indicators Database | |
| 10.5.1 | Regulatory Tier 1 capital to risk-weighted assets | Percent | 2005-2020 | Varies by country/region | Highly likely | Highly likely | Metadata 10-05-01, Global SDG Indicators Database | |
| 10.5.1 | Regulatory capital to assets | Percent | 2005-2020 | Varies by country/region | Highly likely | Highly likely | Metadata 10-05-01, Global SDG Indicators Database | |
| 10.5.1 | Return on assets | Percent | 2005-2020 | Varies by country/region | Highly likely | Highly likely | Metadata 10-05-01, Global SDG Indicators Database | |
| 10.6.1 | Proportion of voting rights of developing countries in international organizations, by organization | Percent | 2000, 2005, 2010, 2015-2020 | Varies by organization | Likely | Unlikely | Metadata 10-06-01, Global SDG Indicators Database | Insufficient data for nowcasting purposes |

| Indicator code | Name | Unit | Data availability | Publication lag | Explanatory variables | Nowcasting feasibility | Data source | Notes |
|---|---|---|---|---|---|---|---|---|
| 10.6.1 | Proportion of voting rights of developing countries in international organizations, by organization | Percent | 2000, 2005, 2010, 2015-2020 | Varies by organization | Likely | Unlikely | Metadata 10-06-01, Global SDG Indicators Database | Insufficient data for nowcasting purposes |
| 10.7.3 | Number of people who died or disappeared in the process of migration towards an international destination | Number | 2014-2021 | Varies by type of reporting | Highly likely | Likely | Missing Migrants Project | Data available annually, monthly, by incident |
| 10.7.4 | Proportion of the population who are refugees, by country of origin | Per 100,000 population | 1951-2020 | Varies by country/region | Highly likely | Highly likely | Metadata 10-07-04, Global SDG Indicators Database | |
| 10.a.1 | Proportion of tariff lines applied to imports from least developed countries and developing countries with zero-tariff | Percent | 2005, 2010, 2015-2019 | 1 year | Likely | Unlikely | Metadata 10-0a-01, Global SDG Indicators Database | Insufficient data for nowcasting purposes |
| 10.b.1 | Total assistance for development, by donor countries | Millions of current United States dollars | 2000-2019 | 2 years | Likely | Likely | Metadata 10-0b-01, Global SDG Indicators Database | |
| 10.b.1 | Total assistance for development, by recipient countries | Millions of current United States dollars | 2000-2019 | 2 years | Likely | Likely | Metadata 10-0b-01, Global SDG Indicators Database | |
| 10.b.1 | Total resource flows for development, by recipient and donor countries | Millions of current United States dollars | 2000-2019 | 2 years | Likely | Likely | Metadata 10-0b-01, Global SDG Indicators Database | |
| 10.c.1 | Corridor remittance costs as a proportion of the amount remitted | Percent | Q42015-Q12021 | 3 quarters | Highly likely | Highly likely | World Bank | Quarterly data release |
| 10.c.1 | Remittance costs as a proportion of the amount remitted | Percent | Q42015-Q12021 | 3 quarters | Highly likely | Highly likely | World Bank | Quarterly data release |
| 10.c.1 | SmaRT corridor remittance costs as a proportion of the amount remitted | Percent | Q42015-Q12021 | 3 quarters | Highly likely | Highly likely | World Bank | Quarterly data release |

| Indicator code | Name | Unit | Data availability | Publication lag | Explanatory variables | Nowcasting feasibility | Data source | Notes |
|---|---|---|---|---|---|---|---|---|
| 11.1.1 | Proportion of urban population living in slums, informal settlements or inadequate housing | Percent | 1990, 1995, 2000, 2005, 2010, 2014, 2016, 2018 | Varies by data source | Highly likely | Unlikely | Metadata 11-01-01, Global SDG Indicators Database | Insufficient data for nowcasting purposes |
| 11.5.1 | Number of deaths and missing persons attributed to disasters | Number | 2005-2020 | Likely 1 year | Highly likely | Likely | Metadata 11-05-01, Global SDG Indicators Database | |
| 11.5.1 | Number of deaths and missing persons attributed to disasters per 100,000 population | Per 100,000 population | 2005-2020 | Likely 1 year | Highly likely | Likely | Metadata 11-05-01, Global SDG Indicators Database | |
| 11.5.1 | Number of deaths due to disaster | Number | 2005-2020 | Likely 1 year | Highly likely | Likely | Metadata 11-05-01, Global SDG Indicators Database | |
| 11.5.1 | Number of directly affected persons attributed to disasters per 100,000 population | Per 100,000 population | 2005-2020 | Likely 1 year | Highly likely | Likely | Metadata 11-05-01, Global SDG Indicators Database | |
| 11.5.1 | Number of injured or ill people attributed to disasters | Number | 2005-2020 | Likely 1 year | Highly likely | Likely | Metadata 11-05-01, Global SDG Indicators Database | |
| 11.5.1 | Number of missing persons due to disaster | Number | 2005-2020 | Likely 1 year | Highly likely | Likely | Metadata 11-05-01, Global SDG Indicators Database | |
| 11.5.1 | Number of people affected by disaster | Number | 2005-2020 | Likely 1 year | Highly likely | Likely | Metadata 11-05-01, Global SDG Indicators Database | |
| 11.5.1 | Number of people whose damaged dwellings were attributed to disasters | Number | 2005-2020 | Likely 1 year | Highly likely | Likely | Metadata 11-05-01, Global SDG Indicators Database | |
| 11.5.1 | Number of people whose destroyed dwellings were attributed to disasters | Number | 2005-2020 | Likely 1 year | Highly likely | Likely | Metadata 11-05-01, Global SDG Indicators Database | |
| 11.5.1 | Number of people whose livelihoods were disrupted or destroyed, attributed to disasters | Number | 2005-2020 | Likely 1 year | Highly likely | Likely | Metadata 11-05-01, Global SDG Indicators Database | |

57                                                                                                           UNCTAD Research Paper No. 82| Indicator code | Name | Unit | Data availability | Publication lag | Explanatory variables | Nowcasting feasibility | Data source | Notes |
|---|---|---|---|---|---|---|---|---|
| 11.6.2 | Annual mean levels of fine particulate matter (e.g. PM2.5 and PM10) in cities (population weighted) | Micrograms per cubic meter | 2011-2016 | Data last released 2018 | Highly likely | Unlikely | Metadata 11-06-01, Global SDG Indicators Database | Insufficient data for nowcasting purposes |
| 11.a.1 | Number of countries that have national urban policies or regional development plans that (a) respond to population dynamics; (b) ensure balanced territorial development; and (c) increase local fiscal space | 1 = YES; 0 = NO (Number of countries for aggregates) | 2018, 2020 | Likely 1-2 years | Likely | Unlikely | Metadata 11-0a-01, Global SDG Indicators Database | Insufficient data for nowcasting purposes |
| 12.2.2 | Domestic material consumption per capita, by type of raw material | Tonnes | 1970-2017 | Data last released 2017 | Highly likely | Highly likely | Metadata 12-02-02, Global SDG Indicators Database | |
| 12.2.2 | Domestic material consumption per unit of GDP, by type of raw material | Kilograms per constant 2010 United States dollars | 1970-2017 | Data last released 2017 | Highly likely | Highly likely | Metadata 12-02-02, Global SDG Indicators Database | |
| 12.2.2 | Domestic material consumption, by type of raw material | Tonnes | 1970-2017 | Data last released 2017 | Highly likely | Highly likely | Metadata 12-02-02, Global SDG Indicators Database | |
| 12.4.1 | Parties meeting their commitments and obligations in transmitting information as required by Basel Convention on hazardous waste, and other chemicals | Number | 2015, 2020 | 1 year | Likely | Unlikely | Metadata 12-04-01, Global SDG Indicators Database | Five year average, data released every 5 years. Insufficient data for nowcasting purposes |
| 12.4.1 | Parties meeting their commitments and obligations in transmitting information as required by Minamata Convention on hazardous waste, and other chemicals | Percent | 2020 | 1 year | Likely | Unlikely | Metadata 12-04-01, Global SDG Indicators Database | Five year average, data released every 5 years. Insufficient data for nowcasting purposes |

| Indicator code | Name | Unit | Data availability | Publication lag | Explanatory variables | Nowcasting feasibility | Data source | Notes |
|---|---|---|---|---|---|---|---|---|
| 12.4.1 | Parties meeting their commitments and obligations in transmitting information as required by Montreal Protocol on hazardous waste, and other chemicals | Number | 2015, 2020 | 1 year | Likely | Unlikely | Metadata 12-04-01, Global SDG Indicators Database | Five year average, data released every 5 years. Insufficient data for nowcasting purposes |
| 12.4.1 | Parties meeting their commitments and obligations in transmitting information as required by Rotterdam Convention on hazardous waste, and other chemicals | Number | 2015, 2020 | 1 year | Likely | Unlikely | Metadata 12-04-01, Global SDG Indicators Database | Five year average, data released every 5 years. Insufficient data for nowcasting purposes |
| 12.4.1 | Parties meeting their commitments and obligations in transmitting information as required by Stockholm Convention on hazardous waste, and other chemicals | Number | 2015, 2020 | 1 year | Likely | Unlikely | Metadata 12-04-01, Global SDG Indicators Database | Five year average, data released every 5 years. Insufficient data for nowcasting purposes |
| 12.a.1 | Installed renewable energy-generating capacity in developing countries | Watts per capita | 2000-2019 | 1 year | Highly likely | Highly likely | Metadata 12-0a-01, Global SDG Indicators Database | |
| 12.b.1 | Implementation of standard accounting tools to monitor the economic and environmental aspects of tourism | Number of tables | 2008-2019 | 1 year | Likely | Likely | Metadata 12-0b-01, Global SDG Indicators Database | |
| 12.b.1 | Implementation of standard accounting tools to monitor the economic and environmental aspects of tourism | Number of tables | 2008-2019 | 1 year | Likely | Likely | Metadata 12-0b-01, Global SDG Indicators Database | |
| 12.b.1 | Implementation of standard accounting tools to monitor the economic and environmental aspects of tourism | Number of tables | 2008-2019 | 1 year | Likely | Likely | Metadata 12-0b-01, Global SDG Indicators Database | |
| 12.c.1 | Fossil-fuel subsidies (consumption and production) | Millions of constant United States dollars | 2015-2019 | 2 year | Likely | Unlikely | Metadata 12-0c-01, Global SDG Indicators Database | Insufficient data for nowcasting purposes |

| Indicator code | Name | Unit | Data availability | Publication lag | Explanatory variables | Nowcasting feasibility | Data source | Notes |
|---|---|---|---|---|---|---|---|---|
| 12.c.1 | Fossil-fuel subsidies (consumption and production) as a proportion of total GDP | Percent | 2015-2019 | 2 year | Likely | Unlikely | Metadata 12-0c-01, Global SDG Indicators Database | Insufficient data for nowcasting purposes |
| 12.c.1 | Fossil-fuel subsidies (consumption and production) per capita | Constant United States dollars | 2015-2019 | 2 year | Likely | Unlikely | Metadata 12-0c-01, Global SDG Indicators Database | Insufficient data for nowcasting purposes |
| 13.1.1 | Number of deaths and missing persons attributed to disasters | Number | 2005-2020 | Likely 1 year | Highly likely | Highly likely | Metadata 13-01-01, Global SDG Indicator Database | |
| 13.1.1 | Number of deaths and missing persons attributed to disasters per 100,000 population | Per 100,000 population | 2005-2020 | Likely 1 year | Highly likely | Highly likely | Metadata 13-01-01, Global SDG Indicator Database | |
| 13.1.1 | Number of deaths due to disaster | Number | 2005-2020 | Likely 1 year | Highly likely | Highly likely | Metadata 13-01-01, Global SDG Indicator Database | |
| 13.1.1 | Number of directly affected persons attributed to disasters per 100,000 population | Per 100,000 population | 2005-2020 | Likely 1 year | Highly likely | Highly likely | Metadata 13-01-01, Global SDG Indicator Database | |
| 13.1.1 | Number of injured or ill people attributed to disasters | Number | 2005-2020 | Likely 1 year | Highly likely | Highly likely | Metadata 13-01-01, Global SDG Indicator Database | |
| 13.1.1 | Number of missing persons due to disaster | Number | 2005-2020 | Likely 1 year | Highly likely | Highly likely | Metadata 13-01-01, Global SDG Indicator Database | |
| 13.1.1 | Number of people affected by disaster | Number | 2005-2020 | Likely 1 year | Highly likely | Highly likely | Metadata 13-01-01, Global SDG Indicator Database | |
| 13.1.1 | Number of people whose damaged dwellings were attributed to disasters | Number | 2005-2020 | Likely 1 year | Highly likely | Highly likely | Metadata 13-01-01, Global SDG Indicator Database | |
| 13.1.1 | Number of people whose destroyed dwellings were attributed to disasters | Number | 2005-2020 | Likely 1 year | Highly likely | Highly likely | Metadata 13-01-01, Global SDG Indicator Database | |

| Indicator code | Name | Unit | Data availability | Publication lag | Explanatory variables | Nowcasting feasibility | Data source | Notes |
|---|---|---|---|---|---|---|---|---|
| 13.1.1 | Number of people whose livelihoods were disrupted or destroyed, attributed to disasters | Number | 2005-2020 | Likely 1 year | Highly likely | Highly likely | Metadata 13-01-01, Global SDG Indicator Database | |
| 13.2.2 | Total greenhouse gas emissions without LULUCF for Annex I Parties | Mt $CO_2$ equivalent | 1990-2019 | 2 years | Highly likely | Highly likely | Metadata 13-02-02, Global SDG Indicator Database | |
| 13.2.2 | Total greenhouse gas emissions without LULUCF for non-Annex I Parties | Mt $CO_2$ equivalent | 2000-2018 | Likely 3 years | Highly likely | Highly likely | Metadata 13-02-02, Global SDG Indicator Database | Length of time series for each country/region varies greatly |
| 14.4.1 | Proportion of fish stocks within biologically sustainable levels | Percent | 1974-2017 | Likely 2-3 years | Highly likely | Highly likely | Metadata 14-04-01, Global SDG Indicator Database | Biennial data release. More recent data likely required |
| 14.5.1 | Average proportion of Marine Key Biodiversity Areas (KBAs) covered by protected areas | Percent | 2000-2020 | 1 year | Highly likely | Highly likely | Metadata 14-05-01, Global SDG Indicator Database | |
| 14.5.1 | Coverage of protected areas in relation to marine areas (Exclusive Economic Zones) | Percent | 2000-2020 | 1 year | Highly likely | Highly likely | Metadata 14-05-01, Global SDG Indicator Database | |
| 14.5.1 | Protected marine area (Exclusive Economic Zones) | Square kilometres | 2000-2020 | 1 year | Highly likely | Highly likely | Metadata 14-05-01, Global SDG Indicator Database | |
| 14.6.1 | Degree of implementation of international instruments aiming to combat illegal, unreported and unregulated fishing | Level of implementation: 1 lowest to 5 highest | 2018, 2020 | 1 year | Likely | Unlikely | Metadata 14-06-01, Global SDG Indicator Database | Biennial data release. Insufficient data for nowcasting purposes |
| 14.7.1 | Sustainable fisheries as a proportion of GDP in small island developing States, least developed countries and all countries | Percent | 2000-2018 (Biennial) | 2 years | Highly likely | Unlikely | Metadata 14-07-01, Global SDG Indicator Database | Biennial data release. Insufficient data for nowcasting purposes |

| Indicator code | Name | Unit | Data availability | Publication lag | Explanatory variables | Nowcasting feasibility | Data source | Notes |
|---|---|---|---|---|---|---|---|---|
| 14.b.1 | Degree of application of a legal/regulatory/policy/institutional framework which recognizes and protects access rights for small-scale fisheries | Level of implementation: 1 lowest to 5 highest | 2011-2018 (Biennial) | 2 years | Likely | Unlikely | Metadata 14-0b-01, Global SDG Indicator Database | Biennial data release. Insufficient data for nowcasting purposes |
| 15.1.1 | Forest area | Thousands of hectares | 1990, 2000, 2010, 2015-2020 | Varies by year | Highly likely | Unlikely | Metadata 15-01-01, Global SDG Indicator Database | Next release of a complete FRA dataset (2021-2025) is scheduled for 2025. Insufficient data for nowcasting purposes |
| 15.1.1 | Forest area as a proportion of total land area | Percent | 1990, 2000, 2010, 2015-2020 | Varies by year | Highly likely | Unlikely | Metadata 15-01-01, Global SDG Indicator Database | Next release of a complete FRA dataset (2021-2025) is scheduled for 2025. Insufficient data for nowcasting purposes |
| 15.1.1 | Land area | Thousands of hectares | 1961-2020 | 1 year | Highly likely | Highly likely | Metadata 15-01-01, Global SDG Indicator Database | |
| 15.1.2 | Average proportion of Freshwater Key Biodiversity Areas (KBAs) covered by protected areas | Percent | 2000-2020 | 1 year | Highly likely | Highly likely | Metadata 15-01-02, Global SDG Indicator Database | |
| 15.1.2 | Average proportion of Terrestrial Key Biodiversity Areas (KBAs) covered by protected areas | Percent | 2000-2020 | 1 year | Highly likely | Highly likely | Metadata 15-01-02, Global SDG Indicator Database | |
| 15.2.1 | Above-ground biomass stock in forest | Tonnes per hectare | 2000, 2010, 2015-2020 | Varies by year | Highly likely | Unlikely | Metadata 15-02-01, Global SDG Indicator Database | Next release of a complete FRA dataset (2021-2025) is scheduled for 2025. Insufficient data for nowcasting purposes |
| 15.2.1 | Forest area annual net change rate | Percent | 2010, 2020 | Varies by year | Highly likely | Unlikely | Metadata 15-02-01, Global SDG Indicator Database | Next release of a complete FRA dataset (2021-2025) is scheduled for 2025. Insufficient data for nowcasting purposes |

| Indicator code | Name | Unit | Data availability | Publication lag | Explanatory variables | Nowcasting feasibility | Data source | Notes |
|---|---|---|---|---|---|---|---|---|
| 15.2.1 | Forest area under an independently verified forest management certification scheme | Thousands of hectares | 2000, 2005, 2010, 2015-2020 | Varies by year | Highly likely | Unlikely | Metadata 15-02-01, Global SDG Indicator Database | Next release of a complete FRA dataset (2021-2025) is scheduled for 2025. Insufficient data for nowcasting purposes |
| 15.2.1 | Proportion of forest area under a long-term management plan | Percent | 2000, 2010, 2015-2020 | Varies by year | Highly likely | Unlikely | Metadata 15-02-01, Global SDG Indicator Database | Next release of a complete FRA dataset (2021-2025) is scheduled for 2025. Insufficient data for nowcasting purposes |
| 15.2.1 | Proportion of forest area within legally established protected areas | Percent | 2000, 2010, 2015-2020 | Varies by year | Highly likely | Unlikely | Metadata 15-02-01, Global SDG Indicator Database | Next release of a complete FRA dataset (2021-2025) is scheduled for 2025. Insufficient data for nowcasting purposes |
| 15.3.1 | Proportion of land that is degraded over total land area | Percent | 2015 | Last data release 2019 | Highly likely | Unlikely | Metadata 15-02-01, Global SDG Indicator Database | Insufficient data for nowcasting purposes |
| 15.4.1 | Coverage by protected areas of important sites for mountain biodiversity | Percent | 2000-2020 | 1 year | Highly likely | Highly likely | Metadata 15-04-01, Global SDG Indicator Database | |
| 15.4.2 | Mountain Green Cover Index | Index | 2000, 2010, 2015, 2018 | All data are already available | Highly likely | Unlikely | Metadata 15-04-02, Global SDG Indicator Database | Insufficient data for nowcasting purposes |
| 15.5.1 | Red List Index | Index | 1993-2021 | Likely 1-2 years | Highly likely | Highly likely | Metadata 15-05-01, Global SDG Indicator Database | November/December data release. Recent data may be forecasted |
| 15.6.1 | Countries that are contracting Parties to the International Treaty on Plant Genetic Resources for Food and Agriculture (PGRFA) | 1 = YES; 0 = NO | 2012-2021 | Likely 1-2 years | Likely | Unlikely | Metadata 15-06-01, Global SDG Indicator Database | November/December data release. Recent data may be forecasted. Insufficient data for nowcasting purposes |
| 15.6.1 | Countries that are parties to the Nagoya Protocol | 1 = YES; 0 = NO | 2012 | Last data release 2016 | Likely | Unlikely | Metadata 15-06-01, Global SDG Indicator Database | Insufficient data for nowcasting purposes |

| Indicator code | Name | Unit | Data availability | Publication lag | Explanatory variables | Nowcasting feasibility | Data source | Notes |
|---|---|---|---|---|---|---|---|---|
| 15.6.1 | Countries that have legislative, administrative and policy framework or measures reported through the Online Reporting System on Compliance of the International Treaty on Plant Genetic Resources for Food and Agriculture (PGRFA) | 1 = YES; 0 = NO | 2016-2021 | Last data release 2021 | Likely | Unlikely | Metadata 15-06-01, Global SDG Indicator Database | Recent data may be forecasted. Insufficient data for nowcasting purposes |
| 15.6.1 | Countries that have legislative, administrative and policy framework or measures reported to the Access and Benefit-Sharing Clearing-House | 1 = YES; 0 = NO | 2012 | Last data release 2016 | Likely | Unlikely | Metadata 15-06-01, Global SDG Indicator Database | Insufficient data for nowcasting purposes |
| 15.6.1 | Total reported number of Standard Material Transfer Agreements (SMTAs) transferring plant genetic resources for food and agriculture to the country | Number | 2012-2021 | Last data release 2021 | Likely | Unlikely | Metadata 15-06-01, Global SDG Indicator Database | Recent data may be forecasted. Insufficient data for nowcasting purposes |
| 15.8.1 | Countries with an allocation from the national budget to manage the threat of invasive alien species | 1 = YES; 0 = NO | 2016, 2020 | Next data release Q1-Q2 2022 | Unlikely | Unlikely | Metadata 15-08-01, Global SDG Indicator Database | Insufficient data for nowcasting purposes |
| 15.8.1 | Legislation, Regulation, Act related to the prevention of introduction and management of Invasive Alien Species | 1 = YES; 0 = NO | 2010, 2016, 2020 | Next data release Q1-Q2 2022 | Unlikely | Unlikely | Metadata 15-08-01, Global SDG Indicator Database | Insufficient data for nowcasting purposes |
| 15.8.1 | National Biodiversity Strategy and Action Plan (NBSAP) targets alignment to Aichi Biodiversity target 9 set out in the Strategic Plan for Biodiversity 2011-2020 | 1 = YES; 0 = NO | 2016, 2020 | Next data release Q1-Q2 2022 | Unlikely | Unlikely | Metadata 15-08-01, Global SDG Indicator Database | Insufficient data for nowcasting purposes |

| Indicator code | Name | Unit | Data availability | Publication lag | Explanatory variables | Nowcasting feasibility | Data source | Notes |
|---|---|---|---|---|---|---|---|---|
| 15.8.1 | Proportion of countries with National Biodiversity Strategy and Action Plan (NBSAP) targets alignment to Aichi Biodiversity target 9 set out in the Strategic Plan for Biodiversity 2011-2020 | Percent | 2016, 2020 | Next data release Q1-Q2 2022 | Unlikely | Unlikely | Metadata 15-08-01, Global SDG Indicator Database | Insufficient data for nowcasting purposes |
| 15.8.1 | Proportion of countries with allocation from the national budget to manage the threat of invasive alien species | Percent | 2020 | Next data release Q1-Q2 2022 | Likely | Unlikely | Metadata 15-08-01, Global SDG Indicator Database | Insufficient data for nowcasting purposes |
| 15.8.1 | Proportion of recipient countries of global funding with access to any funding from global financial mechanisms for projects related to invasive alien species management | Percent | 2020 | Next data release Q1-Q2 2022 | Likely | Unlikely | Metadata 15-08-01, Global SDG Indicator Database | Insufficient data for nowcasting purposes |
| 15.8.1 | Recipient countries of global funding with access to any funding from global financial mechanisms for projects related to invasive alien species management | 1 = YES; 0 = NO | 2016, 2020 | Next data release Q1-Q2 2022 | Likely | Unlikely | Metadata 15-08-01, Global SDG Indicator Database | Insufficient data for nowcasting purposes |
| 15.9.1 | Countries that established national targets in accordance with Aichi Biodiversity Target 2 of the Strategic Plan for Biodiversity 2011-2020 in their National Biodiversity Strategy and Action Plans | 1 = YES; 0 = NO | 2020 | 1 year | Unlikely | Unlikely | Metadata 15-09-01, Global SDG Indicator Database | Insufficient data for nowcasting purposes |

| Indicator code | Name | Unit | Data availability | Publication lag | Explanatory variables | Nowcasting feasibility | Data source | Notes |
|---|---|---|---|---|---|---|---|---|
| 15.9.1 | Countries with integrated biodiversity values into national accounting and reporting systems, defined as implementation of the System of Environmental-Economic Accounting | 1 = YES; 0 = NO | 2020 | 1 year | Unlikely | Unlikely | Metadata 15-09-01, Global SDG Indicator Database | Insufficient data for nowcasting purposes |
| 15.a.1 | Total official development assistance for biodiversity, by donor countries | Millions of constant 2018 United States dollars | 2002-2018 | 1 year | Likely | Likely | Metadata 15-0a-01; Global SDG Indicator Database | 2019 data not yet released |
| 15.a.1 | Total official development assistance for biodiversity, by recipient countries | Millions of constant 2018 United States dollars | 2002-2018 | 1 year | Likely | Likely | Metadata 15-0a-01; Global SDG Indicator Database | 2019 data not yet released |
| 15.b.1 | Total official development assistance for biodiversity, by donor countries | Millions of constant 2018 United States dollars | 2002-2018 | 1 year | Likely | Likely | Metadata 15-0a-01; Global SDG Indicator Database | 2019 data not yet released |
| 15.b.1 | Total official development assistance for biodiversity, by recipient countries | Millions of constant 2018 United States dollars | 2002-2018 | 1 year | Likely | Likely | Metadata 15-0a-01; Global SDG Indicator Database | 2019 data not yet released |
| 16.3.2 | Unsentenced detainees as a proportion of overall prison population | Percent | 2005, 2015, 2018 | Annual | Highly likely | Unlikely | Metadata 16-03-02, Global SDG Indicator Database | Insufficient data for nowcasting purposes |

| Indicator code | Name | Unit | Data availability | Publication lag | Explanatory variables | Nowcasting feasibility | Data source | Notes |
|---|---|---|---|---|---|---|---|---|
| 16.5.2 | Proportion of businesses that had at least one contact with a public official and that paid a bribe to a public official, or were asked for a bribe by those public officials during the previous 12 months | Percent | 2006-2019 | Varies by country/region | Highly likely | Likely | Metadata 16-05-02, Global SDG Indicator Database | Length of time series for each country/region varies greatly |
| 16.7.1 | Number of chairs of permanent committees, by age sex and focus of the committee, Joint Committees | Number | 2009-2021 | Varies by election date | Likely | Likely | Metadata 16-07-01, IPU | February data release. Length of time series for each country/region varies greatly |
| 16.7.1 | Number of chairs of permanent committees, by age sex and focus of the committee, Lower Chamber or Unicameral | Number | 2009-2021 | Varies by election date | Likely | Likely | Metadata 16-07-01, IPU | February data release. Length of time series for each country/region varies greatly |
| 16.7.1 | Number of chairs of permanent committees, by age sex and focus of the committee, Upper Chamber | Number | 2009-2021 | Varies by appointment date | Unlikely | Unlikely | Metadata 16-07-01, IPU | February data release. Length of time series for each country/region varies greatly |
| 16.7.1 | Number of speakers in parliament, by age and sex, Lower Chamber or Unicameral | Number | 2019-2021 | Varies by appointment date | Likely | Likely | Metadata 16-07-01, IPU | February data release. Length of time series for each country/region varies greatly |
| 16.7.1 | Number of speakers in parliament, by age and sex, Upper Chamber | Number | 2019-2021 | Varies by appointment date | Unlikely | Unlikely | Metadata 16-07-01, IPU | February data release. Length of time series for each country/region varies greatly |
| 16.7.1 | Number of youth in parliament (age 45 or below), Lower Chamber or Unicameral (Number) | Number | 2009-2021 | Varies by election date | Likely | Likely | Metadata 16-07-01, Global SDG Indicator Database | February data release. Length of time series for each country/region varies greatly |
| 16.7.1 | Number of youth in parliament (age 45 or below), Upper Chamber (Number) | Number | 2009-2021 | Varies by appointment date | Unlikely | Unlikely | Metadata 16-07-01, Global SDG Indicator Database | February data release. Length of time series for each country/region varies greatly |

| Indicator code | Name | Unit | Data availability | Publication lag | Explanatory variables | Nowcasting feasibility | Data source | Notes |
|---|---|---|---|---|---|---|---|---|
| 16.7.1 | Proportion of youth in parliament (age 45 or below), Lower Chamber or Unicameral | Percent | 2009-2021 | Varies by election date | Likely | Likely | Metadata 16-07-01, Global SDG Indicator Database | February data release. Length of time series for each country/region varies greatly |
| 16.7.1 | Proportion of youth in parliament (age 45 or below), Upper Chamber | Percent | 2009-2021 | Varies by appointment date | Unlikely | Unlikely | Metadata 16-07-01, Global SDG Indicator Database | February data release. Length of time series for each country/region varies greatly |
| 16.7.1 | Ratio for female members of parliaments (Ratio of the proportion of women in parliament in the proportion of women in the national population with the age of eligibility as a lower bound boundary), Lower Chamber or Unicameral | Percent | 2009-2021 | Varies by election date | Likely | Likely | Metadata 16-07-01, Global SDG Indicator Database | February data release. Length of time series for each country/region varies greatly |
| 16.7.1 | Ratio for female members of parliaments (Ratio of the proportion of women in parliament in the proportion of women in the national population with the age of eligibility as a lower bound boundary), Upper Chamber | Percent | 2009-2021 | Varies by appointment date | Unlikely | Unlikely | Metadata 16-07-01, Global SDG Indicator Database | February data release. Length of time series for each country/region varies greatly |
| 16.7.1 | Ratio of young members in parliament (Ratio of the proportion of young members in parliament (age 45 or below) in the proportion of the national population (age 45 or below) with the age of eligibility as a lower bound boundary), Lower Chamber or Unicameral | Percent | 2009-2021 | Varies by election date | Likely | Likely | Metadata 16-07-01, Global SDG Indicator Database | February data release. Length of time series for each country/region varies greatly |

| Indicator code | Name | Unit | Data availability | Publication lag | Explanatory variables | Nowcasting feasibility | Data source | Notes |
|---|---|---|---|---|---|---|---|---|
| 16.7.1 | Ratio of young members in parliament (Ratio of the proportion of young members in parliament (age 45 or below) in the proportion of the national population (age 45 or below) with the age of eligibility as a lower bound boundary), Upper Chamber | Percent | 2009-2021 | Varies by appointment date | Unlikely | Unlikely | Metadata 16-07-01, Global SDG Indicator Database | February data release. Length of time series for each country/region varies greatly |
| 16.8.1 | Proportion of members of developing countries in international organizations, by organization | Percent | 2000, 2005, 2010, 2015-2020 | Varies by organization | Likely | Unlikely | Metadata 16-08-01, Global SDG Indicator Database | Length of time series for each country/region varies greatly |
| 16.8.1 | Proportion of voting rights of developing countries in international organizations, by organization | Percent | 2000, 2005, 2010, 2015-2021 | Varies by organization | Likely | Unlikely | Metadata 16-08-01, Global SDG Indicator Database | Length of time series for each country/region varies greatly |
| 16.9.1 | Proportion of children under 5 years of age whose births have been registered with a civil authority, by age | Percent | 2006-2020 | Varies by country/region | Highly likely | Highly likely | Metadata 16-09-01, Global SDG Indicator Database | Sporadic data availability at the country level. Length of time series for each country/region varies greatly |
| 16.10.2 | Number of countries that adopt and implement constitutional, statutory and/or policy guarantees for public access to information | Number | 2021 | No annual lag | Unlikely | Unlikely | Metadata 16-10-02, Global SDG Indicator Database | Insufficient data for nowcasting purposes |
| 16.a.1 | Countries with National Human Rights Institutions and no status with the Paris Principles, C status | 1 = YES; 0 = NO | 2000, 2005, 2010, 2015-2020 | Likely 1 year | Unlikely | Unlikely | Metadata 16-0a-01, Global SDG Indicator Database | Sporadic data availability at the country level. Length of time series for each country/region varies greatly |
| 16.a.1 | Countries with National Human Rights Institutions in compliance with the Paris Principles, A status | 1 = YES; 0 = NO | 2000, 2005, 2010, 2015-2020 | Likely 1 year | Unlikely | Unlikely | Metadata 16-0a-01, Global SDG Indicator Database | Sporadic data availability at the country level. Length of time series for each country/region varies greatly |

| Indicator code | Name | Unit | Data availability | Publication lag | Explanatory variables | Nowcasting feasibility | Data source | Notes |
|---|---|---|---|---|---|---|---|---|
| 16.a.1 | Countries with National Human Rights Institutions not fully compliant with the Paris Principles, B status | 1 = YES; 0 = NO | 2000, 2005, 2010, 2015-2020 | Likely 1 year | Unlikely | Unlikely | Metadata 16-0a-01, Global SDG Indicator Database | Sporadic data availability at the country level. Length of time series for each country/region varies greatly |
| 16.a.1 | Countries with no application for accreditation with the Paris Principles, D status | 1 = YES; 0 = NO | 2000, 2005, 2010, 2015-2020 | Likely 1 year | Unlikely | Unlikely | Metadata 16-0a-01, Global SDG Indicator Database | Sporadic data availability at the country level. Length of time series for each country/region varies greatly |
| 16.a.1 | Proportion of countries that applied for accreditation as independent National Human Rights Institutions in compliance with the Paris Principles | Percent | 2000, 2005, 2010, 2015-2020 | Likely 1 year | Unlikely | Unlikely | Metadata 16-0a-01, Global SDG Indicator Database | Sporadic data availability at the country level. Length of time series for each country/region varies greatly |
| 16.a.1 | Proportion of countries with independent National Human Rights Institutions in compliance with the Paris Principles | Percent | 2000, 2005, 2010, 2015-2020 | Likely 1 year | Unlikely | Unlikely | Metadata 16-0a-01, Global SDG Indicator Database | Sporadic data availability at the country level. Length of time series for each country/region varies greatly |
| 17.1.1 | Total government revenue (budgetary central government) as a proportion of GDP | Percent | 2000-2019 | 2 years | Highly likely | Highly likely | Metadata 17-01-01, Global SDG Indicator Database | |
| 17.1.1 | Total government revenue, in local currency | Number | 2000-2019 | 2 years | Highly likely | Highly likely | Metadata 17-01-01, Global SDG Indicator Database | |
| 17.1.2 | Proportion of domestic budget funded by domestic taxes | Percent | 2000-2020 | Varies by country/region | Highly likely | Highly likely | Metadata 17-01-02, Global SDG Indicator Database | |
| 17.2.1 | Net official development assistance (ODA) as a percentage of OECD-DAC donors' GNI, by donor countries | Percent | 1960-2020 | 1 year | Likely | Likely | Metadata 17-02-01, OECD | December data release. |

| Indicator code | Name | Unit | Data availability | Publication lag | Explanatory variables | Nowcasting feasibility | Data source | Notes |
|---|---|---|---|---|---|---|---|---|
| 17.2.1 | Net official development assistance (ODA) from OECD-DAC countries, by donor countries | Millions of constant 2019 United States dollars | 1960-2020 | 1 year | Likely | Likely | Metadata 17-02-01, OECD | December data release |
| 17.2.1 | Net official development assistance (ODA) to LDCs as a percentage of OECD-DAC donors' GNI, by donor countries | Percent | 1960-2019 | Likely 2 years | Likely | Likely | Metadata 17-02-01, OECD | December data release |
| 17.2.1 | Net official development assistance (ODA) to LDCs from OECD-DAC countries, by donor countries | Millions of constant 2019 United States dollars | 1960-2019 | Likely 2 years | Likely | Likely | Metadata 17-02-01, OECD | December data release |
| 17.2.1 | Net official development assistance (ODA) to landlocked developing countries as a percentage of OECD-DAC donors' GNI, by donor countries | Percent | 1960-2019 | Likely 2 years | Likely | Likely | Metadata 17-02-01, OECD | December data release |
| 17.2.1 | Net official development assistance (ODA) to landlocked developing countries from OECD-DAC countries, by donor countries | Millions of constant 2019 United States dollars | 1960-2019 | Likely 2 years | Likely | Likely | Metadata 17-02-01, OECD | December data release |
| 17.2.1 | Net official development assistance (ODA) to small island states (SIDS) as a percentage of OECD-DAC donors' GNI, by donor countries | Percent | 1960-2019 | Likely 2 years | Likely | Likely | Metadata 17-02-01, OECD | December data release |

| Indicator code | Name | Unit | Data availability | Publication lag | Explanatory variables | Nowcasting feasibility | Data source | Notes |
|---|---|---|---|---|---|---|---|---|
| 17.2.1 | Net official development assistance (ODA) to small island states (SIDS) from OECD-DAC countries, by donor countries | Millions of constant 2019 United States dollars | 1960-2019 | Likely 2 years | Likely | Likely | Metadata 17-02-01, OECD | December data release |
| 17.2.1 | Official development assistance (ODA) as a percentage of OECD-DAC donors' GNI on grant equivalent basis, by donor countries | Percent | 1960-2020 | 1 year | Likely | Likely | Metadata 17-02-01, OECD | December data release |
| 17.2.1 | Official development assistance (ODA) from OECD-DAC countries on grant equivalent basis, by donor countries | Millions of constant 2019 United States dollars | 1960-2020 | 1 year | Likely | Likely | Metadata 17-02-01, OECD | December data release |
| 17.3.1 | Foreign direct investment (FDI) inflows | Millions of United States dollars | 1980-2020 | Likely 1 year | Highly likely | Highly likely | Global SDG Indicator Database, INE | Tier 1 (FDI)/Tier 2 (ODA, SSC) |
| 17.3.2 | Volume of remittances (in United States dollars) as a proportion of total GDP | Percent | 1970-2020 | Likely 1 year | Highly likely | Highly likely | Metadata 17-03-02, Global SDG Indicator Database | |
| 17.4.1 | Debt service as a proportion of exports of goods and services | Percent | 1970-2019 | 2 years | Highly likely | Highly likely | Metadata 17-04-01, Global SDG Indicator Database | |
| 17.6.1 | Fixed Internet broadband subscriptions per 100 inhabitants, by speed | Per 100 inhabitants | 2000-2019 | Varies by country/region | Highly likely | Highly likely | Metadata 17-06-01, Global SDG Indicator Database | Biannual data release. Length of time series for each country/region varies greatly |
| 17.6.1 | Number of fixed Internet broadband subscriptions, by speed | Number | 2000-2019 | Varies by country/region | Highly likely | Highly likely | Metadata 17-06-01, Global SDG Indicator Database | Biannual data release. Length of time series for each country/region varies greatly |

| Indicator code | Name | Unit | Data availability | Publication lag | Explanatory variables | Nowcasting feasibility | Data source | Notes |
|---|---|---|---|---|---|---|---|---|
| 17.8.1 | Proportion of individuals using the Internet | Per 100 inhabitants | 2000-2019 | Varies by country/region | Highly likely | Highly likely | Metadata 17-08-01, Global SDG Indicator Database | Biannual data release. Length of time series for each country/region varies greatly |
| 17.9.1 | Dollar value of financial and technical assistance (including through North-South, South-South and triangular cooperation) committed to developing countries | Millions of 2019 United States dollars | 1960-2019 | 2 years | Likely | Likely | Metadata 17-09-01, Global SDG Indicator Database | |
| 17.10.1 | Worldwide weighted tariff-average, most-favoured-nation status, by type of product | Percent | 2005-2019 | 2 years | Likely | Likely | Metadata 17-10-01, Global SDG Indicator Database | |
| 17.10.1 | Worldwide weighted tariff-average, preferential status, by type of product | Percent | 2005-2019 | 2 years | Likely | Likely | Metadata 17-10-01, Global SDG Indicator Database | |
| 17.11.1 | Developing countries' and least developed countries' share of global merchandise exports | Percent | 2000-2019 | 2 years | Highly likely | Highly likely | Metadata 17-11-01, Global SDG Indicator Database | |
| 17.11.1 | Developing countries' and least developed countries' share of global merchandise imports | Percent | 2000-2019 | 2 years | Highly likely | Highly likely | Metadata 17-11-01, Global SDG Indicator Database | |
| 17.11.1 | Developing countries' and least developed countries' share of global services exports | Percent | 2000-2019 | 2 years | Highly likely | Highly likely | Metadata 17-11-01, Global SDG Indicator Database | |
| 17.11.1 | Developing countries' and least developed countries' share of global services imports | Percent | 2000-2019 | 2 years | Highly likely | Highly likely | Metadata 17-11-01, Global SDG Indicator Database | |

| Indicator code | Name | Unit | Data availability | Publication lag | Explanatory variables | Nowcasting feasibility | Data source | Notes |
|---|---|---|---|---|---|---|---|---|
| 17.12.1 | Average tariff applied by developed countries, most-favored nation status, by type of product | Percent | 2000, 2005, 2010-2019 | Depends on launch of SDG Monitoring Report | Likely | Likely | Metadata 17-12-01, Global SDG Indicator Database | |
| 17.12.1 | Average tariff applied by developed countries, preferential status, by type of product | Percent | 2000, 2005, 2010-2019 | Depends on launch of SDG Monitoring Report | Likely | Likely | Metadata 17-12-01, Global SDG Indicator Database | |
| 17.18.2 | Number of countries that have national statistical legislation that complies with the Fundamental Principles of Official Statistics | 1 = YES; 0 = NO | 2017-2019 | 2018 data released Q1 2017 | Likely | Unlikely | Metadata 17-18-02, Global SDG Indicator Database | Insufficient data for nowcasting purposes |
| 17.18.3 | Countries with national statistical plans that are fully funded | 1 = YES; 0 = NO | 2007-2015, 2019-2020 | 2015 data released 2017 | Likely | Unlikely | Metadata 17-18-03, Global SDG Indicator Database | Insufficient data for nowcasting purposes |
| 17.18.3 | Countries with national statistical plans that are under implementation | 1 = YES; 0 = NO | 2007-2015, 2019-2020 | 2015 data released 2017 | Likely | Unlikely | Metadata 17-18-03, Global SDG Indicator Database | Insufficient data for nowcasting purposes |
| 17.18.3 | Countries with national statistical plans with funding from Government | 1 = YES; 0 = NO | 2007-2015, 2019-2020 | 2015 data released 2017 | Likely | Unlikely | Metadata 17-18-03, Global SDG Indicator Database | Insufficient data for nowcasting purposes |
| 17.18.3 | Countries with national statistical plans with funding from donors | 1 = YES; 0 = NO | 2007-2015, 2019-2020 | 2015 data released 2017 | Likely | Unlikely | Metadata 17-18-03, Global SDG Indicator Database | Insufficient data for nowcasting purposes |
| 17.18.3 | Countries with national statistical plans with funding from others | 1 = YES; 0 = NO | 2007-2015, 2019-2020 | 2015 data released 2017 | Likely | Unlikely | Metadata 17-18-03, Global SDG Indicator Database | Insufficient data for nowcasting purposes |
| 17.19.1 | Dollar value of all resources made available to strengthen statistical capacity in developing countries | Current United States dollars | 2006-2013, 2018 | 2013 data released 2016 | Likely | Unlikely | Metadata 17-19-01, Global SDG Indicator Database | Insufficient data for nowcasting purposes |

| Indicator code | Name | Unit | Data availability | Publication lag | Explanatory variables | Nowcasting feasibility | Data source | Notes |
|---|---|---|---|---|---|---|---|---|
| 17.19.2 | Countries that have conducted at least one population and housing census in the last 10 years | 1 = YES; 0 = NO | 2019 | 2019 data released 2020/2021 | Likely | Unlikely | Metadata 17-19-02, Global SDG Indicator Database | Insufficient data for nowcasting purposes |
| 17.19.2 | Countries with birth registration data that are at least 90 percent complete | 1 = YES; 0 = NO | 2015-2019 | Likely 2 years | Likely | Unlikely | Metadata 17-19-02, Global SDG Indicator Database | Insufficient data for nowcasting purposes |
| 17.19.2 | Countries with death registration data that are at least 75 percent complete | 1 = YES; 0 = NO | 2015-2019 | Likely 2 years | Likely | Unlikely | Metadata 17-19-02, Global SDG Indicator Database | Insufficient data for nowcasting purposes |
| 17.19.2 | Proportion of countries that have conducted at least one population and housing census in the last 10 years | Percent | 2019 | 2019 data released 2020/2021 | Likely | Unlikely | Metadata 17-19-02, Global SDG Indicator Database | Insufficient data for nowcasting purposes |
| 17.19.2 | Proportion of countries with birth registration data that are at least 90 percent complete | Percent | 2015-2019 | Likely 2 years | Likely | Unlikely | Metadata 17-19-02, Global SDG Indicator Database | Insufficient data for nowcasting purposes |
| 17.19.2 | Proportion of countries with death registration data that are at least 75 percent complete | Percent | 2015-2019 | Likely 2 years | Likely | Unlikely | Metadata 17-19-02, Global SDG Indicator Database | Insufficient data for nowcasting purposes |